\documentclass[11pt]{article}
\usepackage[margin=1in]{geometry}
\usepackage[
  backend=biber,
  style=authoryear,
  natbib=true,
  sorting=nyt,
  maxcitenames=3,
  mincitenames=1,
  maxbibnames=6,
  minbibnames=6,
  uniquelist=false,
  uniquename=false,
  giveninits=true,
  dashed=false,
  doi=false,
  isbn=false,
  url=false,
  eprint=false
]{biblatex}
\DeclareNameAlias{author}{family-given}
\DeclareNameAlias{editor}{family-given}
\DeclareDelimFormat{multinamedelim}{\addcomma\space}
\DeclareDelimFormat{finalnamedelim}{\addspace\bibstring{and}\space}
\DeclareDelimFormat{nameyeardelim}{\addcomma\space}
\DefineBibliographyStrings{english}{andothers={et\addabbrvspace al\adddot}}
\DeclareFieldFormat[article,misc,incollection,inproceedings]{title}{\MakeSentenceCase*{#1}}
\DeclareFieldFormat[book]{title}{\mkbibemph{#1}}
\DeclareFieldFormat{journaltitle}{\mkbibemph{#1}}
\DeclareFieldFormat[article]{volume}{#1}
\DeclareFieldFormat{pages}{#1}
\DeclareFieldFormat{edition}{\mkbibordedition{#1}\addspace Edition}
\AtEveryBibitem{%
  \clearfield{number}%
  \clearfield{month}%
  \clearfield{day}%
  \clearfield{issn}%
  \clearfield{doi}%
  \clearfield{url}%
  \clearfield{urldate}%
  \clearfield{isbn}%
  \clearfield{language}%
}
\DeclareBibliographyDriver{article}{%
  \usebibmacro{bibindex}%
  \usebibmacro{begentry}%
  \printnames{author}%
  \setunit{\addspace}%
  \printtext[parens]{\printfield{year}}%
  \setunit{\addperiod\space}%
  \printfield{title}%
  \setunit{\addperiod\space}%
  \printfield{journaltitle}%
  \setunit{\addspace}%
  \printfield{volume}%
  \iffieldundef{pages}{}{\setunit{\addcomma\space}\printfield{pages}}%
  \setunit{\addperiod}%
  \usebibmacro{finentry}%
}
\DeclareBibliographyDriver{book}{%
  \usebibmacro{bibindex}%
  \usebibmacro{begentry}%
  \printnames{author}%
  \setunit{\addspace}%
  \printtext[parens]{\printfield{year}}%
  \setunit{\addperiod\space}%
  \printfield{title}%
  \setunit{\addperiod\space}%
  \iffieldundef{edition}{}{\printfield{edition}\setunit{\addperiod\space}}%
  \printlist{publisher}%
  \setunit{\addperiod}%
  \usebibmacro{finentry}%
}
\DeclareBibliographyDriver{misc}{%
  \usebibmacro{bibindex}%
  \usebibmacro{begentry}%
  \printnames{author}%
  \setunit{\addspace}%
  \printtext[parens]{\printfield{year}}%
  \setunit{\addperiod\space}%
  \printfield{title}%
  \setunit{\addperiod\space}%
  \printfield{howpublished}%
  \setunit{\addperiod}%
  \usebibmacro{finentry}%
}
\usepackage{amsmath,amssymb,amsfonts,amsthm}
\usepackage{booktabs,multirow,longtable,array,float,graphicx,xurl,rotating}
\usepackage{xcolor}
\usepackage[hidelinks,hypertexnames=false]{hyperref}
\allowdisplaybreaks
\graphicspath{{./}}
\newtheorem{theorem}{Theorem}
\newtheorem{lemma}{Lemma}

\newtheorem{proposition}{Proposition}
\theoremstyle{definition}

\newtheorem{example}{Example}
\newtheorem{remark}{Remark}
\numberwithin{equation}{section}

\newcommand{\E}{\mathrm{E}}

\newcommand{\twoline}[2]{\shortstack{#1\\(#2)}}
\newcommand{\meanrev}[1]{#1}

\hypersetup{
  pdftitle={Data-Adaptive Integration with External Summary Data for Outcome Mean Estimation},
  pdfauthor={Kosuke Morikawa, Sho Komukai, and Satoshi Hattori},
  pdfkeywords={data integration, entropy balancing, external summary statistics, covariate shift, double robustness}
}
\title{Data-Adaptive Integration with External Summary Data\\for Outcome Mean Estimation}
\author{Kosuke Morikawa$^{1}$, Sho Komukai$^{2}$, and Satoshi Hattori$^{3,4}$\\[0.6em]
\small $^{1}$Department of Statistics, Iowa State University, Ames, Iowa 50011, U.S.A.\\
\small $^{2}$Department of Health Data Science, Tokyo Medical University, Tokyo 160-8402, Japan\\
\small $^{3}$Department of Biomedical Statistics, The University of Osaka, Osaka, Japan\\
\small $^{4}$Integrated Frontier Research for Medical Science Division, Institute for Open and\\
\small Transdisciplinary Research Initiatives, The University of Osaka, Osaka, Japan\\[0.4em]
\small \texttt{morikawa@iastate.edu; komukai.sho.7w@tokyo-med.ac.jp}\\
\small \texttt{hattoris@biostat.med.osaka-u.ac.jp}}
\date{}

\begin{document}
\maketitle
\begin{abstract}
\begin{sloppypar}
Combining an internal individual-level study with readily available external summary statistics promises major efficiency gains at minimal additional cost, yet heterogeneity between sources can bias estimates for the internal target population. We develop a generalized entropy-balancing integration strategy that calibrates the internal individual-level sample to externally reported moments while retaining the internal population as the target, explicitly permitting a biased external sample. The weighted-regression version of our estimator is doubly robust: it remains consistent when either the outcome-regression model or the entropy-balancing model is correctly specified. When multiple balancing specifications are plausible, we introduce a data-adaptive entropy-family selection rule. For the final borrowing decision, we propose a bootstrap-based criterion comparing stabilized mean squared error (MSE) estimates for the selected entropy-balancing estimator and the internal sample mean. This criterion is selection consistent under fixed alternatives and reverts to the internal estimator when a nonvanishing bias is detected. Separately, under a linear homoscedastic benchmark, the asymptotic efficiency criteria admit geometric interpretations through the Mahalanobis distance and Pearson chi-squared divergence. The entropy-balancing estimators and numerical-experiment routines are implemented in the \textsc{R} package \texttt{daisy}. Simulations show stable MSE reductions for the weighted-regression estimator across calibrated distributional shifts and the predicted reversion toward the internal estimator under fixed simultaneous misspecification as the sample size increases. An application to nationwide public-access defibrillation records in Japan illustrates the resulting MSE-based borrowing decision.
\end{sloppypar}
\end{abstract}

\section{Introduction}\label{sec:1}

Incorporating external information is a practical strategy for improving estimator efficiency, particularly in research fields where data collection entails considerable time and financial costs. In disciplines such as epidemiology, clinical trials, genetics, and genomics, integrating historical data from previous studies has become common practice to enhance estimation accuracy \citep{lee_estimation_2020, li_target_2020}. Combining such external data, whether at the individual or summary level, with ongoing studies can lead to more precise estimates \citep{pocock_combination_1976, riley_individual_2021}. When individual-level external data are available, the integration approach generally depends on the data structure and includes methods such as individual patient data meta-analysis \citep{stewart_meta-analysis_1993, riley_meta-analysis_2010}, missing-data techniques \citep{schafer_missing_2002, little_statistical_2019}, and causal-inference approaches \citep{yang_combining_2020}. In recent machine-learning research, there is a growing trend toward integrating external synthetic data, rather than raw external data, into analysis frameworks \citep{bashari2025statistical}.

Conversely, when only summary-level external data are available, specialized techniques are required for effective integration. If both internal and summary-level external datasets originate from the same underlying population, established meta-analytic methodologies are typically employed \citep{kundu_generalized_2019, borenstein_introduction_2009}. In randomized controlled trials, sample sizes are usually determined based on primary endpoints, which may result in insufficient statistical power for secondary endpoints. To address this limitation, methods such as the ``test-then-pool'' approach, in which the internal control group is first compared with the external (historical) control data and pooling is performed only if no significant differences are detected, have been proposed \citep{viele_use_2014}. Furthermore, matching-adjusted indirect comparison (MAIC) methods have been developed to adjust for population heterogeneity between internal and external datasets \citep{signorovitchComparativeEffectivenessHeadtoHead2010}. This approach reweights individual-level internal data to align with summary-level baseline characteristics of the external data, enabling indirect comparisons between treatment groups from internal and external sources. Although MAIC is primarily designed for indirect treatment comparisons, its application to augment internal control data with external control information is conceptually related but methodologically distinct, as it does not involve direct pooling of data.

When substantial heterogeneity exists between the populations represented by the internal and external datasets, evaluating generalizability and transportability becomes essential. Within the context of randomized controlled trials, assessing treatment effects in broader populations beyond the specific trial sample is critical. This issue has been extensively addressed not only in randomized trials but also in observational studies \citep{cole_generalizing_2010}, using approaches such as constrained or restricted maximum likelihood \citep{chatterjee_constrained_2016, han_empirical_2019}, empirical likelihood \citep{qin_empirical_1994, qin_combining_2000, qin_using_2015}, and calibration \citep{lumley_connections_2011, yang_combining_2020}. Recently, entropy balancing has gained prominence as an effective data-integration technique due to its robustness against model misspecification, making it particularly suitable for addressing covariate shift \citep{shimodaira_improving_2000} and population heterogeneity \citep{hainmueller_entropy_2012}. Entropy balancing has also been used to align covariate distributions with external targets in transportability analyses \citep{josey_transporting_2021}. 

When heterogeneity is pronounced, a central question is how to handle differences between the external and internal data-generating distributions. Some recent work targets the external population and estimates optimal treatment regimes under distributional shift \citep{chu_targeted_2023}. By contrast, other approaches target the internal population and aim to improve regression estimation by borrowing external information. For example, James--Stein-type shrinkage has been proposed under Gaussian errors \citep{HanEtAl_Biometrics_2024}, and automatic tuning procedures that downweight external information under large discrepancies have been developed \citep{ZhaiHan_EJS_2024}. For binary outcomes, methods leveraging proportionality of reduced coefficients across studies have been proposed for incorporating external summary coefficients into internal generalized linear models \citep{TaylorChoiHan_Biometrika_2023}. Many of these methods rely on a correctly specified conditional outcome model to enable valid borrowing. Recent methods can also select or downweight external summaries in a data-adaptive way \citep{huang_simultaneous_2023,ZhaiHan_EJS_2024,fang_integrated_2025}. What seems to be less developed, however, is a fully estimable final rule that protects the internal target from nonvanishing borrowing bias and is selection consistent under fixed alternatives. We address this with a bootstrap-based comparison of stabilized MSE estimates. Separately, under a linear homoscedastic model, the asymptotic efficiency criteria have geometric interpretations in terms of the Mahalanobis distance and Pearson chi-squared divergence.

In this paper, we propose an estimator for the outcome mean, tailored to the current-study distribution, that integrates external summary-level information via generalized entropy balancing. For such target parameters, the requirement of correctly specifying the conditional mean model is often stringent. We also introduce a weighted-regression version that uses (i) an outcome-regression model and (ii) an entropy-function model for the balancing step and enjoys double robustness: consistency holds if either component is correctly specified.

Although this resembles the classical notion of doubly robust estimation in the missing-data and causal-inference literature (e.g., \citealp{robins_regression_1994}), both the setting and the construction here are fundamentally different. Standard doubly robust estimators typically combine an outcome-regression model with an individual-level propensity/missingness model and require access to external individual-level data. In contrast, this weighted version does not require individual-level external data at all: it borrows information only through externally reported summary moments and a balancing step. As a result, efficiency is not characterized as cleanly as in the usual semiparametric theory for fully observed data, and the benefit of integration depends on how far the internal and external distributions are from each other.

 When multiple models are posited for the entropy functions, our estimation procedure allows data-adaptive selection among them. Moreover, our approach does not require the commonly assumed availability of external regression coefficients; instead, it can operate using only moment information on both the outcome and the covariates. After selecting the entropy family, the final borrowing decision uses the bootstrap-based stabilized MSE criterion: it selects the internal-only estimator when the data reveal a nonvanishing bias and otherwise compares estimated MSEs. The Mahalanobis distance and Pearson chi-squared divergence do not enter this selector; under a linear homoscedastic benchmark, they geometrically characterize the asymptotic efficiency criteria.

The main contributions of this paper are:
\begin{enumerate}
\item We propose an estimator that targets the internal-population mean while borrowing information from external \emph{summary} statistics, even when the external sample is biased. This estimator is constructed by linking the internal and external populations via two rounds of generalized entropy balancing.
\item We establish double robustness of the weighted-regression version and asymptotic normality, and we provide a data-adaptive procedure for selecting the first-step entropy function among multiple candidates.
\item We propose a bootstrap-based stabilized MSE criterion for choosing between the entropy-balancing estimator and the internal sample mean, and prove its selection consistency under fixed alternatives. Separately, under a linear homoscedastic benchmark, we express the asymptotic efficiency criteria geometrically in terms of the Mahalanobis distance and Pearson $\chi^2$ divergence.
\item We demonstrate practical performance in simulations and in a nationwide public-access defibrillation application, and we provide accompanying software implementation.
\end{enumerate}

To aid the reader, we proceed in increasing order of the amount of external information available: we first develop estimators that use only external means of $X$ and $Y$ (Condition~(C2)), and then discuss extensions that incorporate richer external summaries (e.g., regression coefficients) when they are available. 
The remainder of the paper is organized as follows. Section~\ref{sec:2} introduces the setup and assumptions. Section~\ref{sec:3} presents the proposed generalized entropy balancing estimators and discusses computation and model selection. Section~\ref{sec:4} establishes large-sample properties. Section~\ref{sec:5} introduces the bootstrap-based, selection-consistent MSE criterion and then gives a separate geometric interpretation of the asymptotic efficiency criteria. Sections~\ref{sec:6} and \ref{sec:7} report simulation and real-data results, respectively, and Section~\ref{sec:8} concludes with a discussion.

\section{Basic Setup}\label{sec:2}
Suppose that $Z_i = (X_i^\top, Y_i)^\top$ for $i = 1, \dots, n$ are $n$ independent and identically distributed observations drawn from an internal data source, where $X_i$ denotes a vector of covariates and $Y_i$ denotes the outcome variable. In contrast, let $Z_i = (X_i^\top, Y_i)^\top$ for $i = n+1, \dots, n+n_1$ represent $n_1$ independent and identically distributed observations from an external data source, for which only summary statistics of $X$ and $Y$ are available. Throughout, we allow $X$ to include prespecified transformations of baseline variables (e.g., polynomial terms or interactions). We use a tilde to denote vector augmentation by a generic intercept; for instance, $\tilde{X} = (1, X^\top)^\top$, and we use the same convention for other symbols.

Let $S_i$ be a source indicator that equals one if observation $Z_i$ originates from the external source. We write $\E_{\mathrm{in}}$ and $\E_{\mathrm{ex}}$ for expectations under the internal and external distributions, respectively. Our target is the internal-population mean
$
\theta^*=\E_{\mathrm{in}}(Y).
$
Although the same idea can be extended to more general estimating-equation parameters, we focus on the mean in this paper. This is because the external information considered here is relatively limited. Specifically, as in (C2), the external information is given only through a simple aggregate summary, such as a mean. Such information is often easy to obtain in practice, but it does not provide detailed information about the external distribution. Therefore, the internal-population mean is a natural and interpretable target for studying how this type of limited external information can be used to improve estimation.

Our analysis relies on the following two conditions:
\begin{enumerate}

    \item[(C1)] $\E_{\mathrm{in}}(Y\mid X=x)=\E_{\mathrm{ex}}(Y\mid X=x)$ for every $x$ in the external support, and $f_{\mathrm{ex}}\ll f_{\mathrm{in}}$, where $f_{\mathrm{in}}$ and $f_{\mathrm{ex}}$ denote the internal and external covariate densities or mass functions.

    \item[(C2)] The external-sample means of $X$ and $Y$, denoted by $\hat{\mu}_{x|\mathrm{ex}}$ and $\hat{\mu}_{y|\mathrm{ex}}$, are available.

\end{enumerate}

Our aim is to construct an estimator that is more efficient than one relying solely on the internal data by incorporating external summaries $\hat{\mu}_{x|\mathrm{ex}}$ and $\hat{\mu}_{y|\mathrm{ex}}$. Condition~(C1) combines conditional-mean transportability with one-sided overlap. The stronger condition $Y\perp S\mid X$, which implies equality of the full conditional distribution and is the analogue of missing at random (MAR) in a missing-data formulation \citep{rubin_inference_1976}, is sufficient but not required here. The weaker mean restriction is enough because the target and the point-estimation equations used below depend on $Y$ through the common conditional mean in (C1); conditional variances and higher conditional moments may differ across sources, except where a common variance is imposed separately in (C6) for the closed-form efficiency calculation. This assumption is most plausible when $X$ includes all effect modifiers and key prognostic factors that differ across settings and when $Y$ is defined and measured comparably across studies. The support condition $f_{\mathrm{ex}}\ll f_{\mathrm{in}}$ provides the overlap needed to reweight the internal distribution toward the external distribution, while heterogeneity in the marginal distribution of $X$ remains permitted.

Condition~(C2) is the central assumption underlying our analysis. It does not require auxiliary information, such as externally estimated regression coefficients or other model-specific parameters, that is often assumed in related work. This minimal requirement enhances both the generality and the practical applicability of our framework. When regression coefficients from an external dataset are available, our two-step balancing construction admits a natural extension beyond linear regression; see Supplementary Section~S3.

The following two examples illustrate settings in which the internal population remains the target but external summaries may improve precision.

\noindent\textit{Example 1 (Historical controls).} In a randomized trial, external summaries from historical controls can improve estimation of the current-population control mean and hence the average treatment effect, while the current individual-level sample remains the target.

\noindent\textit{Example 2 (Probability and non-probability samples).} A probability sample can define the target population while inexpensive non-probability-sample summaries are borrowed after calibration; naive pooling may be biased when participant characteristics differ.

\section{Generalized Entropy Balancing for Integration of Summary Data}\label{sec:3}
To address the discrepancy between the internal and external distributions, we introduce a two-stage set of balancing weights. In the first stage, we balance selected internal covariate moments to the externally reported moments. In the second stage, we use the external outcome mean to calibrate a scalar constructed from the first-step weights and the internal outcome regression. The final estimator nevertheless targets the internal-population mean.

We begin by assuming that the regression function is linear. This assumption will be relaxed and further discussed in Section~\ref{sec:3.3}.
\begin{itemize}
\item[(C3)] The common conditional mean in (C1) is linear: $\E_{\mathrm{in}}(Y\mid X=x)=\E_{\mathrm{ex}}(Y\mid X=x)=\beta^\top\tilde{x}$,
\end{itemize}
where $\tilde{x}=(1,x^\top)^\top$. 
We first compute $\hat{\beta}_{\mathrm{in}}$, the ordinary least squares estimator based on the internal data. Let $\beta^*_{\mathrm{in}}$ and $\beta^*_{\mathrm{ex}}$ denote the population-level linear regression coefficients for the internal and the external data, respectively. Under (C1) and (C3), we have $\beta^*_{\mathrm{in}}=\beta^*_{\mathrm{ex}}$.

\subsection{Derivation of the Proposed Estimator}
Let $G_1:(0,\infty)\to\mathbb R$ be twice continuously differentiable and strictly convex, with $G_1'$ mapping $(0,\infty)$ onto $\mathbb R$. For any entropy function $G$, write $g=G'$, $\rho=g^{-1}$, and $F(u)=\sup_{w>0}\{uw-G(w)\}$ for its derivative, inverse link, and Fenchel conjugate, respectively. The Kullback--Leibler (KL) choice is $G_1(w)=w\log w-w$; we also consider the Lambert--$W$ (LW), quadratic log-sum (QLS), and tempered-softplus (TS) families. Their explicit $(G,g,\rho,F)$ formulas and tuning parameters are collected in Supplementary Section~S2.

\begin{figure}[htbp]
\begin{center}
\includegraphics[width=66mm]{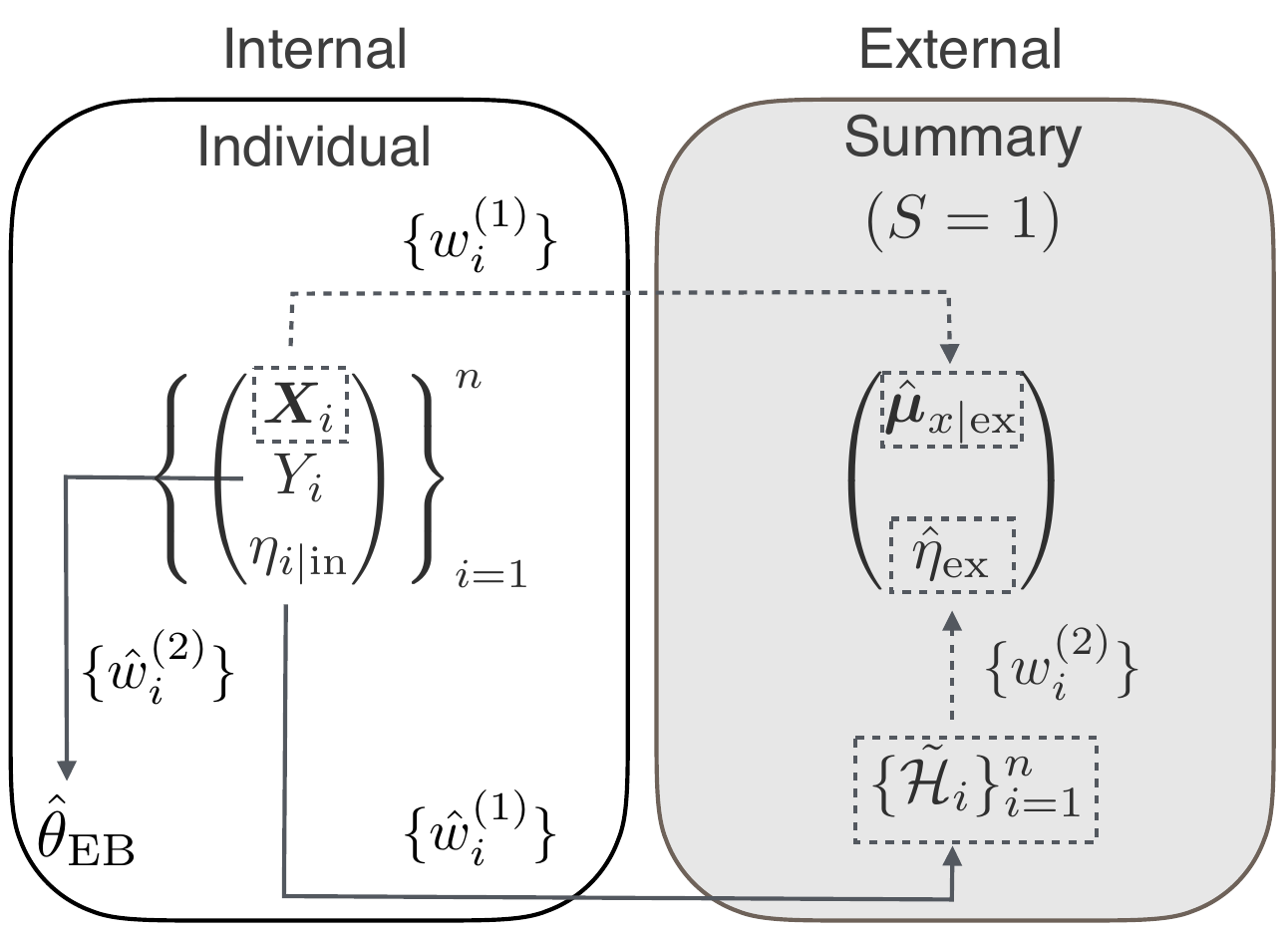}
\caption{Conceptual diagram for the construction of the proposed estimator $\hat{\theta}_{\mathrm{EB}}$. The dashed lines indicate the summary statistics used in the entropy-balancing constraints, while the solid lines show how $\{\tilde{\mathcal{H}}_i\}$ and $\hat{\theta}_{\mathrm{EB}}$ are constructed from the estimated weights.}
\label{fig:1}
\end{center}
\end{figure}

We introduce two sets of weights to connect the internal and external populations. The first set balances the selected moments of $X$, so that the reweighted internal moments match their external counterparts (see Figure~\ref{fig:1} for an illustration). Specifically, we solve the following optimization problem with respect to $w^{(1)}=(w^{(1)}_1,\dots, w^{(1)}_n)$:
\begin{align*}
\max_{w^{(1)}}\ -n^{-1}\sum_{i=1}^n G_1(w^{(1)}_i)
\end{align*}
subject to
\begin{align}
n^{-1}\sum_{i=1}^n w^{(1)}_i = 1, \quad 
n^{-1}\sum_{i=1}^n w^{(1)}_i X_i = \hat{\mu}_{x|\mathrm{ex}}.
\label{EB1}
\end{align}

Intuitively, in the case of no discrepancy between the internal and external datasets, uniform weights $w_i \equiv 1$ would nearly satisfy the constraints in \eqref{EB1}. Otherwise, the estimated weights $\hat{w}_i^{(1)}$ contain information about the density ratio $f_{\mathrm{ex}}(x) / f_{\mathrm{in}}(x)$, since the law of large numbers yields
\[
n^{-1}\sum_{i=1}^n \frac{f_{\mathrm{ex}}(X_i)}{f_{\mathrm{in}}(X_i)} X_i = \E_{\mathrm{ex}}(X) + o_p(1),
\]
where $f_{\mathrm{in}}(x)$ and $f_{\mathrm{ex}}(x)$ are the density functions of $x$ in the internal and external data, respectively, and $\E_{\mathrm{ex}}(X)$ is the external-population mean of $X$.

Next, we introduce a second stage of entropy balancing. Define $\hat{\mathcal{H}}_i = \hat{w}^{(1)}_i \hat{\eta}_{i|\mathrm{in}}$, where $\hat{\eta}_{i|\mathrm{in}} = \eta(X_i; \hat{\beta}_{\mathrm{in}}) = \hat{\beta}_{\mathrm{in}}^\top \tilde{X}_i$, and define $\hat{\eta}_{\mathrm{ex}}  = \hat{\mu}_{y|\mathrm{ex}}$. Hereafter, as stated in Section~\ref{sec:2}, the tilde notation denotes an augmented vector that includes a constant term: $\tilde{\mu}_{x|\mathrm{ex}} = (1, \hat{\mu}_{x|\mathrm{ex}}^\top)^\top$, $\tilde{\mathcal{H}}_i = (1, \hat{w}^{(1)}_i \hat{\eta}_{i|\mathrm{in}})^\top$, and $\tilde{\eta}_{\mathrm{ex}} = (1, \hat{\eta}_{\mathrm{ex}})^\top$.

Let $G_2:(0,\infty)\to\mathbb{R}$ denote another twice continuously differentiable and strictly convex entropy function satisfying the same boundary conditions; $G_2$ need not coincide with $G_1(\cdot)$. Section~\ref{sec:selection} discusses the role of $G_1$ and $G_2$. The second set of weights $\hat{w}_i^{(2)}$ is estimated by solving:
\begin{align*}
\max_{w^{(2)}}\ -n^{-1}\sum_{i=1}^n G_2(w^{(2)}_i)
\end{align*}
subject to
\begin{align}
n^{-1}\sum_{i=1}^n w^{(2)}_i = 1, \quad 
n^{-1}\sum_{i=1}^n w^{(2)}_i \hat{\mathcal{H}}_i = \hat{\eta}_{\mathrm{ex}}. \label{EB2}
\end{align}

Our final entropy-balancing (EB) estimator $\hat{\theta}_{\mathrm{EB}}$ is defined as
\begin{align}
\hat{\theta}_{\mathrm{EB}} = n^{-1}\sum_{i=1}^n \hat{w}^{(2)}_iY_i. \label{EB}
\end{align}

The role of the second-step weights $w^{(2)}$ is more nuanced than that of the first-step weights $w^{(1)}$. Using $\{\hat{w}_i^{(1)}\}$, we map $\hat{\eta}_{i|\mathrm{in}}$ to $\hat{\mathcal{H}}_i$. At the population level, write $w^{*(1)}(X)=\rho_1(\lambda_1^{*\top}\tilde X)$, $\mathcal H^*=w^{*(1)}(X)\beta_{\mathrm{in}}^{*\top}\tilde X$, and $\eta_{\mathrm{ex}}^*=\E_{\mathrm{ex}}(Y)$. Condition~(C1) and either (C3) or (C3)$'$ imply $\E_{\mathrm{in}}(\mathcal H^*)=\eta_{\mathrm{ex}}^*$; hence the population second-step weight is uniform, $w^{*(2)}\equiv1$. This is the balance behind the double robustness of EBW consistency; equality of the full conditional distribution of $Y\mid X$ is unnecessary. Although the final weights $\{\hat{w}_i^{(2)}\}$ behave nearly as constants, they still incorporate essential information from $\hat{\mu}_{x|\mathrm{ex}}$ and $\hat{\mu}_{y|\mathrm{ex}}$, thereby offering potential efficiency gains.

\subsection{Estimation Procedure for the Proposed Estimator}
Writing $g_j=G_j'$ and $\rho_j=g_j^{-1}$, convex duality gives
\begin{align}
\ell_1(\hat\lambda_1)
&=n^{-1}\sum_{i=1}^n\rho_1(\hat\lambda_1^\top\tilde X_i)\tilde X_i-\tilde\mu_{x\mid\mathrm{ex}}=0,
\label{est_eq1}\\
\ell_2(\hat\beta_{\mathrm{in}},\hat\lambda_1,\hat\lambda_2)
&=n^{-1}\sum_{i=1}^n\rho_2(\hat\lambda_2^\top\tilde{\mathcal H}_i)\tilde{\mathcal H}_i-\tilde\eta_{\mathrm{ex}}=0,
\label{est_eq2}
\end{align}
so that $\hat w_i^{(1)}=\rho_1(\hat\lambda_1^\top\tilde X_i)$ and $\hat w_i^{(2)}=\rho_2(\hat\lambda_2^\top\tilde{\mathcal H}_i)$. The dual objectives and computational formulas are given in Supplementary Section~S2.

\subsection{Modification for Misspecified Regression Models}
\label{sec:3.3}

When the working regression model is misspecified, the probability limits $\beta_{\mathrm{in}}^*$ and $\beta_{\mathrm{ex}}^*$ can differ under covariate shift \citep{shimodaira_improving_2000}. The consistency argument in Theorem~\ref{consistency} requires the regression limit entering the second-step construction to agree with the external-population regression limit. We therefore seek a condition under which $\beta_{\mathrm{in}}^*=\beta_{\mathrm{ex}}^*$ is restored, and estimate the regression model by minimizing the following weighted objective:
\begin{align}
\sum_{i=1}^n \frac{f_{\mathrm{ex}}(X_i)}{f_{\mathrm{in}}(X_i)} \left\{ Y_i - \eta(X_i; \beta_{\mathrm{in}}) \right\}^2, \label{weight.eq}
\end{align}
Because the density ratio is typically unknown, it must be estimated. A natural choice is to adopt the first-step entropy-balancing weights $\hat{w}^{(1)}_i$ as an estimator of the density ratio. If the true density ratio admits the single-index form, i.e.,
\begin{itemize}
  \item[(C3)$'$] There exists $\lambda_1$ such that $f_{\mathrm{ex}}(x)/f_{\mathrm{in}}(x)=\rho_1(\lambda_1^\top \tilde{x})$,
\end{itemize}
then the conditional-mean restriction in (C1) is sufficient for the weighted regression. Indeed, a change of measure and (C1) give
\begin{align}
\E_{\mathrm{in}}\!\left[\frac{f_{\mathrm{ex}}(X)}{f_{\mathrm{in}}(X)}\tilde X\{Y-\eta(X;\beta)\}\right]
=\E_{\mathrm{ex}}\!\left[\tilde X\{Y-\eta(X;\beta)\}\right].
\label{eq:weighted-mean-identity}
\end{align}
Thus the weighted internal normal equation is the external-population normal equation without requiring equality of the conditional variances or the full conditional distributions. The resulting estimator of the regression coefficients is consistent for $\beta^*_{\mathrm{ex}}$, so the required identity $\beta^*_{\mathrm{in}}=\beta^*_{\mathrm{ex}}$ holds in our notation. When $\hat\beta_{\mathrm{in}}$ in \eqref{EB} is obtained from \eqref{weight.eq} with $w_i=\hat w_i^{(1)}$, we denote the resulting weighted-regression entropy-balancing estimator by $\hat\theta_{\mathrm{EBW}}$ (EBW); EB uses ordinary least squares, whereas EBW uses the first-step-weighted regression.

\noindent\textbf{Practical estimation algorithm.} For each candidate $G_1$: (1) solve \eqref{est_eq1} for $\hat w_i^{(1)}$; (2) estimate $\beta_{\mathrm{in}}$ by ordinary least squares for EB or by the sample version of \eqref{weight.eq} with $w_i=\hat w_i^{(1)}$ for EBW; and (3) form $\tilde{\mathcal H}_i$, solve \eqref{est_eq2}, and obtain $\hat w_i^{(2)}$. If several candidates are considered, Section~\ref{sec:selection} then selects $G_1$; finally, compute \eqref{EB}.


\subsection{Selection of Entropy Functions}\label{sec:selection}
Two entropy functions, $G_1$ and $G_2$, have appeared thus far, and it is useful to clarify their distinct roles. The function $G_1$ governs the modeling of the density ratio through the dual link: as described in Section~3.2, the first-stage weights can be written as $\hat w_i^{(1)}=\rho_1\!\bigl(\hat\lambda_1^\top \tilde X_i\bigr)$ $(i=1,\dots,n)$,
where $\rho_1 = g_1^{-1}$ with $g_1(w)=dG_1(w)/dw$. Hence, the choice of $G_1$ is, in effect, the choice of a model for the density ratio. For example, when the internal and external covariate distributions are Gaussian and $(1,x,x^2)$ is balanced, the density ratio is exponential quadratic and is represented by the KL inverse link. Additional details are in Supplementary Section~S2.
By contrast, for $G_2$ we impose only strict convexity; no additional structural constraints are required. Its role is to provide a common measure of how far the second-step weights depart from uniformity and, through that measure, to select among candidate choices of $G_1$. Specifically, let $\{G_{1[j]}\}_{j=1}^J$ denote multiple candidates for the
first-step entropy functions.
As a second-step entropy function, we take a function $G_2$ that measures the deviation
of the weights from the uniform vector $(1,\dots,1)$, that is, how close $w_i$ is to $1$ under the
working assumptions. Typical choices include
\[
G_2(w)=w\log w \quad\text{(KL-based)} \qquad\text{and}\qquad G_2(w)=(w-1)^2 \quad\text{(quadratic)}.
\]
In particular, under the normalization $n^{-1}\sum_{i=1}^n w_i=1$, the KL form is equivalent
(up to an additive constant) to $w\log w - w$; hence adopting $G_2(w)=w\log w - w$ is essentially
without loss for the purpose of selecting among the $G_1$ candidates.

Given $\{G^{(j)}_1\}_{j=1}^J$, we choose the model whose induced weights are closest to uniform under $G_2$:
\[\hat j\in\arg\min_{1\leq j\leq J}\hat Q_n(j).\]
Here
\begin{align}
 \hat{Q}_n(j)= n^{-1}\sum_{i=1}^n G_2\!\bigl(\hat w^{(2)}_{i[j]}\bigr), \label{select}
\end{align}
where $\hat{w}^{(2)}_{i[j]}$ $(i=1,\dots,n;\; j=1,\dots,J)$ are the estimated second-step balancing weights obtained when $G_{1[j]}$ is used in the first-step balancing. We refer to the resulting step-1 entropy-family-selected estimator as S1EF.

\begin{proposition}[Entropy-family selection]\label{prop:entropy-selection}
\leavevmode\par
Let $Q^*(j)=\E_{\mathrm{in}}\{G_2(w^{*(2)}_{[j]})\}$ and $\mathcal J_0=\arg\min_{1\le j\le J}Q^*(j)$. If the candidate set is finite and the regularity conditions (C4)--(C5) hold for each candidate, then $\Pr(\hat j\in\mathcal J_0)\to1$. If $\mathcal J_0$ is a singleton, $\hat j$ is consistent for that family.
\end{proposition}
Any candidate whose limiting second-step weights are uniform belongs to $\mathcal J_0$ by Jensen's inequality; several candidates may tie, particularly under population homogeneity. The proof is given in Supplementary Section~S1. This first selection chooses the entropy family. The separate final decision between S1EF and the internal sample mean is made by the MSE criterion in Section~\ref{sec:5}.

When richer external summaries are sufficient to identify the external average prediction, the same two-step construction extends beyond the linear baseline; see Supplementary Section~S3.

\section{Large Sample Theory}\label{sec:4}

We state the large-sample results for $\hat\theta_{\mathrm{EBW}}$, since $\hat\theta_{\mathrm{EB}}$ is the special case in which the first-step weights are not used when estimating $\beta_{\mathrm{in}}$. Throughout Sections~\ref{sec:4} and \ref{sec:5}, assume that the external sample size satisfies $n/n_1\to\kappa$. Let
\[
\hat\xi=(\hat\beta_{\mathrm{in}}^\top,\hat\lambda_1^\top,\hat\lambda_2^\top,\hat\theta_{\mathrm{EBW}})^\top
\]
and define the estimating equations $\ell(\xi)=(\ell_0^\top,\ell_1^\top,\ell_2^\top,\ell_3)^\top$ by
\[
\ell_0(\beta_{\mathrm{in}},\lambda_1)
=n^{-1}\sum_{i=1}^n \rho_1(\lambda_1^\top\tilde X_i)\varphi(X_i,Y_i;\beta_{\mathrm{in}}),
\]
together with \eqref{est_eq1}, \eqref{est_eq2}, and
\begin{align}
\ell_3(\beta_{\mathrm{in}},\lambda_1,\lambda_2,\theta)
=n^{-1}\sum_{i=1}^n \rho_2(\lambda_2^\top\tilde{\mathcal H}_i)(\theta-Y_i).
\label{est_eq3}
\end{align}
Here $\varphi$ is the estimating function for the working regression model. Regularity conditions (C4)--(C5), collected in the Supplementary Material, are standard smoothness and moment assumptions ensuring uniform laws of large numbers and a central limit theorem for these estimating equations. This stacked-equation form is useful conceptually because it shows that the asymptotic theory is driven by a regression estimating equation and the balancing equations, not by a fully specified likelihood for $Y\mid X$.

\begin{theorem}\label{consistency}
Under Conditions~(C1), (C2), (C4), and (C3) or (C3)$'$, the estimator $\hat\theta_{\mathrm{EBW}}$ is consistent for $\theta^*$.
\end{theorem}

Theorem~\ref{consistency} formalizes the double robustness of the procedure: consistency follows if either the working outcome model is correct, as in (C3), or the first-step entropy model correctly represents the density ratio, as in (C3)$'$. In either case, the corresponding population equations imply $\E_{\mathrm{in}}(\mathcal H^*)=\eta_{\mathrm{ex}}^*$ and therefore $w^{*(2)}\equiv1$. Thus uniformity of the second-step weights is a doubly robust consequence used to establish consistency. This statement does not make the efficiency comparison doubly robust; the closed-form variance representations in Section~\ref{sec:5} require the additional model and variance assumptions stated there.

Under (C1), (C2), (C4), (C5), and either (C3) or (C3)$'$, the stacked estimating equations yield the asymptotic linear representation
\begin{align}
n^{1/2}(\hat\theta_{\mathrm{EBW}}-\theta^*)
&=n^{-1/2}\sum_{i=1}^n\psi(Z_i)-\sqrt\kappa\,\tau_1^\top W_1-\sqrt\kappa\,\tau_2^\top W_2+o_p(1),
\label{eq:ALR}
\end{align}
where $W_1,W_2$ are the Gaussian limits of the external summaries and $\psi,\tau_1,\tau_2$ are given in Supplementary Section~S1.

\begin{theorem}\label{thm:CLT}
Under Conditions~(C1), (C2), (C4), (C5), and either (C3) or (C3)$'$,
\[
n^{1/2}(\hat\theta_{\mathrm{EBW}}-\theta^*)\xrightarrow{d}N(0,\Sigma_{\mathrm{EBW}}),
\]
where $\Sigma_{\mathrm{EBW}}$ is the variance implied by \eqref{eq:ALR}.
\end{theorem}

The explicit form of $\Sigma_{\mathrm{EBW}}$, the Jacobian components, and the proofs of Proposition~\ref{prop:entropy-selection} and Theorems~\ref{consistency}--\ref{thm:CLT} are given in Supplementary Section~S1.

\section{Final Selection and Efficiency Interpretation}\label{sec:5}

\begingroup
\subsection{MSE-based final selection}\label{sec:mse-select}
After selecting the first-step entropy family by \eqref{select}, let $\hat\theta_{\mathrm P}$ denote the resulting EB or EBW estimator and let $\hat\theta_{\mathrm I}=\bar Y$ be the internal sample mean (SM). Write $\theta_{\mathrm P}^\dagger=\operatorname*{plim}\hat\theta_{\mathrm P}$ and $b_{\mathrm P}=\theta_{\mathrm P}^\dagger-\theta^*$. From paired bootstrap replicates of $(\hat\theta_{\mathrm I},\hat\theta_{\mathrm P})$, estimate
\[
V_{\mathrm I}=\operatorname{Var}(\hat\theta_{\mathrm I}),\quad
V_{\mathrm P}=\operatorname{Var}(\hat\theta_{\mathrm P}),\quad
V_{\mathrm D}=\operatorname{Var}(\hat\theta_{\mathrm P}-\hat\theta_{\mathrm I})
\]
by $\hat V_{\mathrm I},\hat V_{\mathrm P},\hat V_{\mathrm D}$. For $c_n\to\infty$ and $c_n/\sqrt n\to0$, define
\begin{align}
\hat B_{\mathrm P}^2
&=\left[(\hat\theta_{\mathrm P}-\hat\theta_{\mathrm I})^2-\hat V_{\mathrm D}\right]_+
\mathbf 1\{|\hat\theta_{\mathrm P}-\hat\theta_{\mathrm I}|>c_n\hat V_{\mathrm D}^{1/2}\},
\label{eq:bias2-stabilized}\\
\hat R_{\mathrm I}&=\hat V_{\mathrm I},\qquad
\hat R_{\mathrm P}=\hat V_{\mathrm P}+\hat B_{\mathrm P}^2,
\label{eq:mse-estimates}
\end{align}
and select $\hat\theta_{\mathrm P}$ when $\hat R_{\mathrm P}\leq\hat R_{\mathrm I}$, otherwise $\hat\theta_{\mathrm I}$. This defines $\hat\theta_{\mathrm{MSE}}$. We use $\hat V_{\mathrm I}=s_Y^2/n$, where $s_Y^2=(n-1)^{-1}\sum_{i=1}^n(Y_i-\bar Y)^2$, and $c_n=\sqrt{\log n}$.

\begin{theorem}[Selection consistency]\label{thm:mse-selection}
Suppose $\hat\theta_{\mathrm I}-\theta^*=O_p(n^{-1/2})$, $\hat\theta_{\mathrm P}-\theta_{\mathrm P}^\dagger=O_p(n^{-1/2})$, and $n\hat V_k\to_p\Sigma_k$ for $k\in\{\mathrm I,\mathrm P,\mathrm D\}$, where $\Sigma_{\mathrm D}>0$. If $b_{\mathrm P}=0$ and $\Sigma_{\mathrm P}\ne\Sigma_{\mathrm I}$, the rule selects the estimator with smaller asymptotic variance with probability tending to one. If $b_{\mathrm P}\ne0$, it selects $\hat\theta_{\mathrm I}$ with probability tending to one.
\end{theorem}
The proof and the limitation under $n^{-1/2}$-order local misspecification are given in Supplementary Section~S1.

\subsection{Geometric interpretation of the efficiency criteria}\label{sec:distance-efficiency}
The bootstrap MSE rule in Section~\ref{sec:mse-select} is the only final selection rule used by the procedure. The following result does not define a distance-based selector; under a linear homoscedastic benchmark, it provides a geometric characterization of the conditions under which EB and EBW are no less efficient than SM.
\begin{theorem}[Geometric efficiency criteria]\label{thm:distance-efficiency}
Assume the common linear mean model and
\begin{enumerate}
\item[(C6)] $\operatorname{Var}(Y\mid X=x,S=s)=\sigma^2$ for $s\in\{0,1\}$,
\end{enumerate}
and define
\[
\alpha=\frac{\operatorname{Cov}_{\mathrm{in}}(Y,\mathcal H^*)}{\operatorname{Var}_{\mathrm{in}}(\mathcal H^*)},\quad
D_1^2=(\mu_{x\mid\mathrm{ex}}^*-\mu_{x\mid\mathrm{in}}^*)^\top\Sigma^{*-1}(\mu_{x\mid\mathrm{ex}}^*-\mu_{x\mid\mathrm{in}}^*),
\]
where $0<\operatorname{Var}_{\mathrm{in}}(\mathcal H^*)<\infty$, $\Sigma^*=\operatorname{Var}_{\mathrm{in}}(X)$, and $\mathcal H^*=w^{*(1)}(X)\beta_{\mathrm{in}}^{*\top}\tilde X$, with $\beta_{\mathrm{in}}^*$ obtained by ordinary least squares for EB and by first-step-weighted least squares for EBW. Under (C1), (C2), (C3), and (C4)--(C6), ordinary least squares gives
\begin{align}
\Sigma_{\mathrm{EB}}
=\operatorname{Var}_{\mathrm{in}}(Y)-\sigma^2\{2\alpha-\alpha^2(1+D_1^2+\kappa)\}.
\label{general_eb_variance}
\end{align}
Thus EB is no less efficient than SM exactly when $2\alpha-\alpha^2(1+D_1^2+\kappa)\ge0$, or, for $\alpha>0$, when
\[
D_1^2\leq 2/\alpha-1-\kappa.
\]
If, additionally, (C3)$'$ holds, the regression coefficients are estimated using the first-step weights, and
\[
D_2^2=\chi^2(f_{\mathrm{ex}}\|f_{\mathrm{in}})<\infty,
\]
then
\begin{align}
\Sigma_{\mathrm{EBW}}
=\operatorname{Var}_{\mathrm{in}}(Y)-\sigma^2\{2\alpha-\alpha^2(1+D_2^2+\kappa)\},
\label{general_ebw_variance}
\end{align}
and, for $\alpha>0$, the corresponding condition is $D_2^2\leq2/\alpha-1-\kappa$. The EBW consistency result is doubly robust, but this closed-form variance calculation uses both working models and (C6).
\end{theorem}

The theoretical variance comparison depends on $\alpha$, so it is useful first to identify a benchmark in which $\alpha$ is known. If the internal and external covariate distributions coincide, then $w^{*(1)}\equiv1$ and $\mathcal H^*$ is the conditional mean, or the population least-squares projection with an intercept. Projection orthogonality gives $\operatorname{Cov}_{\mathrm{in}}(Y,\mathcal H^*)=\operatorname{Var}_{\mathrm{in}}(\mathcal H^*)$, and hence $\alpha=1$. Substituting $\alpha=1$ yields $D_1^2\leq1-\kappa$ and $D_2^2\leq1-\kappa$; when $\kappa=0$, these become the simple unit-threshold conditions $D_1\leq1$ and $D_2^2\leq1$. These are theoretical benchmarks for the asymptotic variance comparison, not inputs to the bootstrap selector. Under heterogeneity, $\alpha$ may be below or above one. Its estimator, detailed interpretation, the Gaussian relation $D_2^2=\exp(D_1^2)-1$, and all proofs are given in Supplementary Section~S1.
\endgroup

\section{Numerical Experiment}\label{sec:6}

\begingroup
We evaluated the procedure in 1,000 Monte Carlo repetitions with an internal sample of $n=200$ and an external sample of $n_1=2{,}000$. Internally, $X_1\sim N(0,1)$ and $X_2\sim\operatorname{Bernoulli}(0.5)$. The first-step balance uses $B(X)=(X_1,X_1^2,X_2)^\top$. External distributions were constructed to have population Mahalanobis distance
\[
D\in\{0,0.25,0.5,1,1.5\}
\]
from the internal distribution with respect to $B(X)$. Here $D$ indexes the designed degree of heterogeneity and is not used by MSE-Select. We considered no heterogeneity, normal mean shifts, normal variance shifts, and shifted-Gamma distributions. The outcome was generated from
\[
Y\mid X_1,X_2\sim N(0.5X_1-X_2+0.5aX_1^2,1),\qquad a\in\{0,1\},
\]
so the working linear outcome model is correct when $a=0$ and misspecified when $a=1$. The precise calibration formulas and the complete numerical summaries are given in Supplementary Section~S5.

For clarity, SM denotes the internal sample mean $\bar Y$, and WSM the naive mean that weights the internal and external sample means by their sample sizes. EB and EBW denote the proposed estimators based on ordinary and first-step-weighted outcome regressions, respectively, while KL denotes a fixed Kullback--Leibler first-step entropy. S1EF denotes the estimator obtained after selecting $G_1$ by the generalized criterion in \eqref{select} over the KL, LW, QLS, and TS families, whereas MSE-Select is the final choice between S1EF and SM in Section~\ref{sec:mse-select}. Within each paired bootstrap, the originally selected entropy family was held fixed, and $c_n=\sqrt{\log n}$ was used in \eqref{eq:bias2-stabilized}. Figure~\ref{fig:2-mean-mse} shows the mean-shift settings; the variance-shift and Gamma figures are in the Supplementary Material.

\begin{figure}[htbp]
\centering
\includegraphics[width=\linewidth]{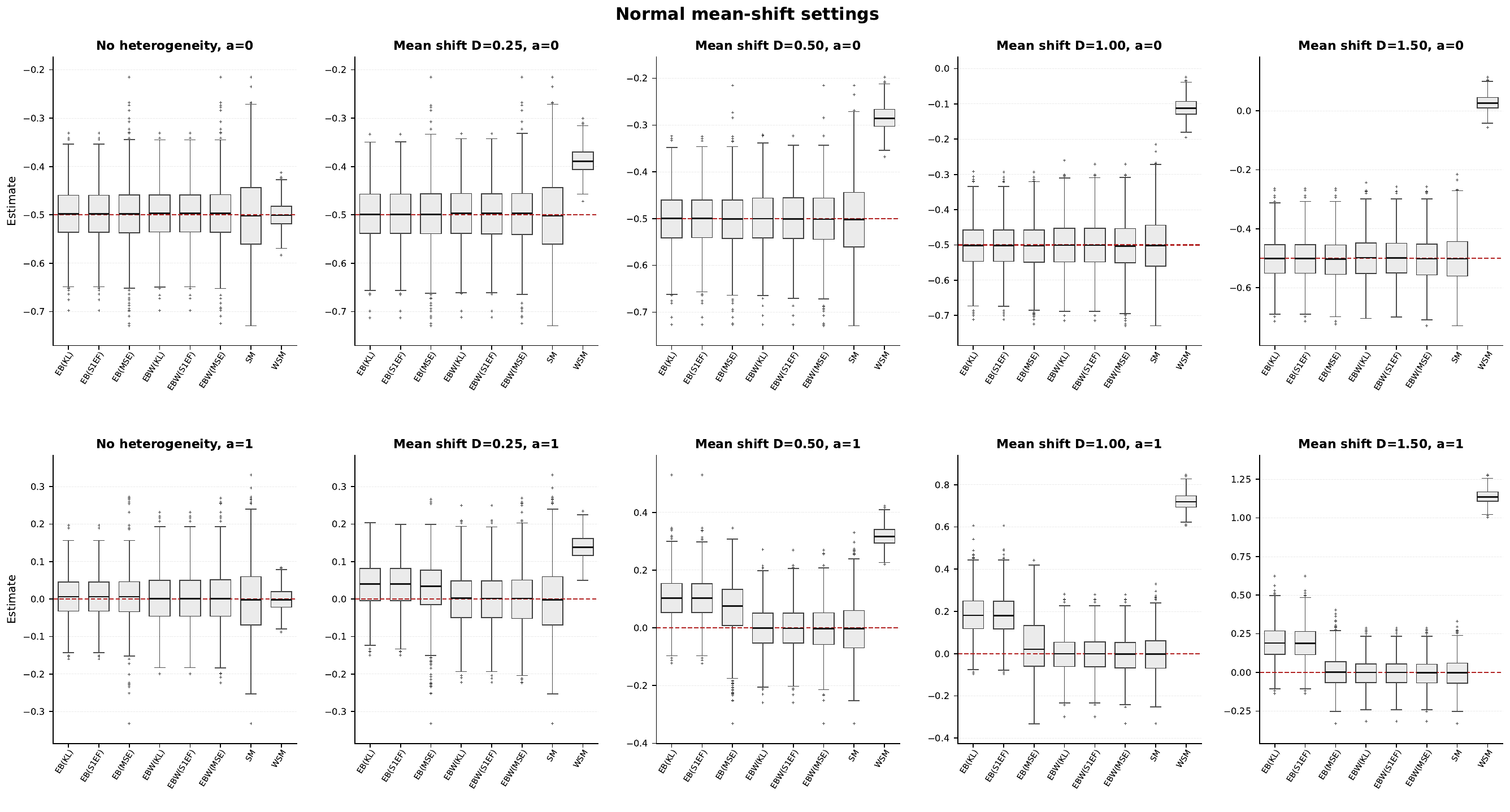}
\caption{Monte Carlo distributions under no heterogeneity and normal mean shifts calibrated by the population Mahalanobis distance $D$. Top row: correctly specified outcome model ($a=0$); bottom row: misspecified outcome model ($a=1$). ``MSE'' denotes the final MSE-selected estimator. Dashed lines mark the true internal-population mean.}
\label{fig:2-mean-mse}
\end{figure}

The weighted-regression estimator EBW performed stably across all scenarios. Relative to SM, the empirical MSE ratio of EBW(MSE-Select) ranged from $0.484$ to $0.927$ when $a=0$ and from $0.581$ to $0.941$ when $a=1$; its absolute Monte Carlo bias was below $0.008$ throughout. The borrowing frequency decreased as heterogeneity increased. For example, with $a=1$ it declined from $97.3\%$ to $76.5\%$ over the normal mean shifts and from $96.5\%$ to $73.4\%$ over the variance shifts. Thus the MSE rule retained most of the efficiency gain when the proposed estimator was stable while increasingly reverting to SM under greater discrepancy.

The unweighted EB version was efficient when the outcome model was correct, but under outcome-model misspecification its finite-sample selector did not always detect moderate bias. In several $a=1$ settings its MSE exceeded that of SM, with the largest ratio equal to $2.289$. This contrast is consistent with the fact that EBW, rather than EB, is the doubly robust version and motivates treating EBW as the primary estimator. Naive pooling became severely biased as the external distribution moved away from the internal target.

We also performed a focused sample-size experiment to evaluate Theorem~\ref{thm:mse-selection}. Under correct working models at $D=0.5$, EBW(MSE-Select) borrowed in $97.8\%$, $99.0\%$, and $99.2\%$ of repetitions for $n=200$, $500$, and $1000$, and its MSE ratios relative to SM were $0.607$, $0.543$, and $0.512$. Under simultaneous misspecification, a Gamma shape shift had $D=0$ while an omitted cubic outcome term generated fixed bias. The borrowing frequency then fell from $30.9\%$ to $4.4\%$ and $0.1\%$, and the MSE ratio approached one ($1.860$, $1.274$, and $1.007$). This supports selection consistency while showing that asymptotic protection need not be exact at $n=200$.
\endgroup

\section{Real Data Analysis}\label{sec:7}

We applied the method to the All-Japan Utstein Registry of out-of-hospital cardiac arrest \citep{jacobs_cardiac_2004, kitamura_nationwide_2010}. We drew an internal sample of $n=200$ patients from the 2019 registry and considered two external datasets of size $1{,}000$: one sampled from the remaining 2019 registry and one sampled from the 2010--2015 registry. Only the external summary means entered the point-estimation procedure; the individual external records were used to construct those summaries and to propagate their sampling uncertainty in the bootstrap. Baseline summaries and absolute standardized mean differences are reported in the Supplementary Material.

\begin{figure}[htbp]
\centering
\includegraphics[width=\linewidth]{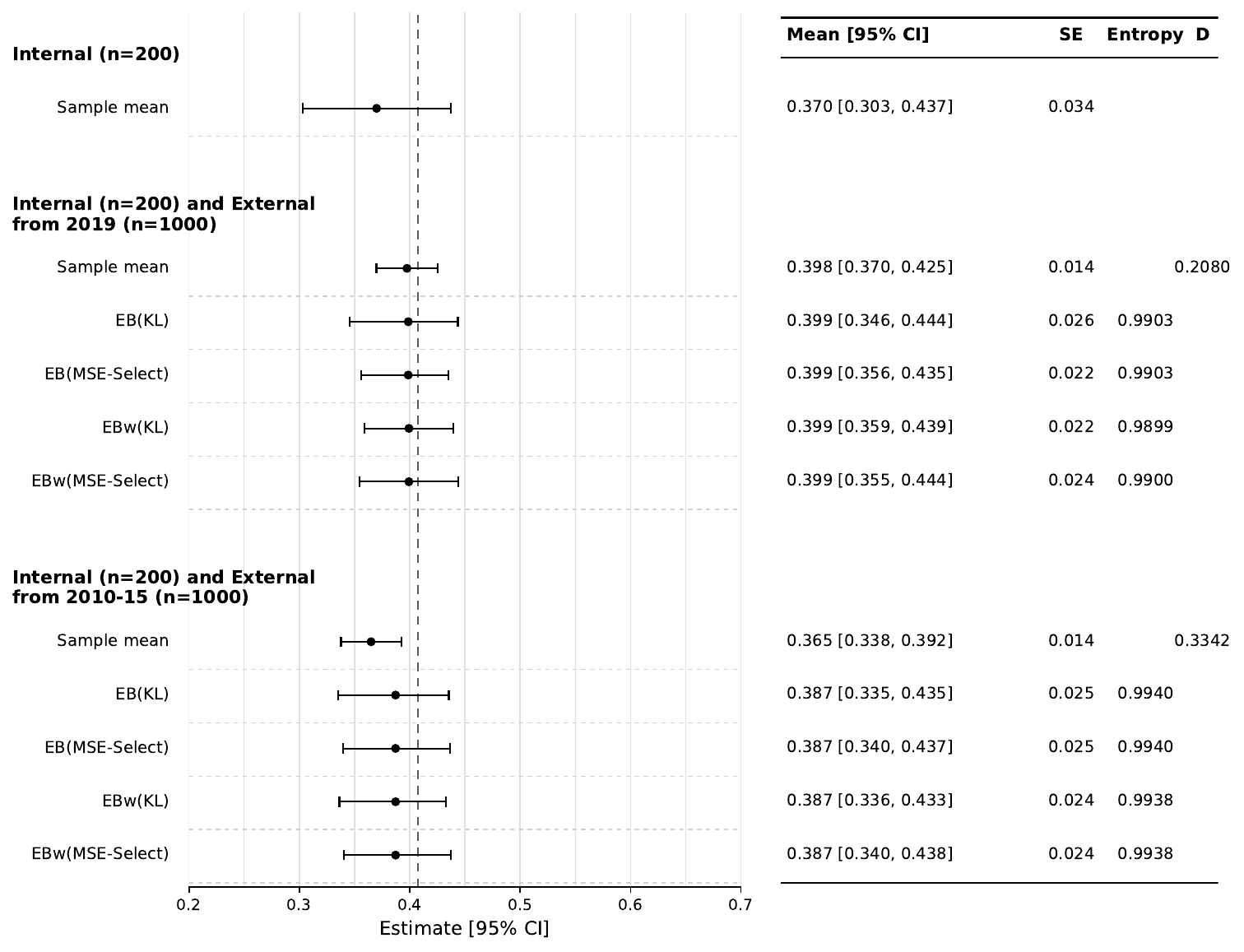}
\caption{Point estimates and component-wise 95\% bootstrap intervals in the real-data analysis. For each external source, the display includes naive pooling, fixed-KL EB/EBW, and the final MSE-selected EB/EBW estimators. ``Entropy'' is $-n^{-1}\sum_i\{w_i^{(2)}\log w_i^{(2)}-w_i^{(2)}\}$, whose maximum under normalized weights is one, and $D$ is the sample Mahalanobis distance, reported descriptively and not used for selection. The intervals correspond to the component selected in the observed data and are not post-selection confidence intervals. The dashed line marks the 2019 registry mean.}
\label{fig:real-data-analysis}
\end{figure}

\begingroup
The internal sample mean was $0.370$ (standard error (SE) $0.034$; 95\% confidence interval (CI) $[0.303,0.437]$). With the 2019 external sample, S1EF selected the TS family with $r=0.1$. The EB(MSE-Select) and EBW(MSE-Select) estimates were both $0.399$, with SEs $0.022$ and $0.024$ and estimated MSE ratios relative to the internal estimator of $0.422$ and $0.504$. With the 2010--2015 external sample, S1EF selected QLS with $r=0.01$. The two selected estimates were both $0.387$, with SEs $0.025$ and $0.024$ and estimated MSE ratios $0.552$ and $0.507$. The paired bootstrap detected no fixed bias in any of the four comparisons, and the MSE criterion selected the proposed estimator. The selected estimates were closer than the internal sample mean to the 2019 registry mean marked in Figure~\ref{fig:real-data-analysis}, and their intervals were shorter. Because the true MSE is not observed in an application, these comparisons are stated in terms of the estimated MSE criterion rather than as a guarantee about realized risk. The detailed MSE components are tabulated in the Supplementary Material.
\endgroup

\section{Discussion and Conclusion}\label{sec:8}

Several directions merit further investigation for generalized entropy balancing with biased external summaries. 
First, the current framework assumes that the individual-level data are sampled from the target population distribution. This limits applicability in settings involving biased internal samples, such as observational studies aiming to estimate average treatment effects. 
Second, the method requires that both the individual-level data and the external summary data be defined over the same set of explanatory variables. When individual-level external data are available, the regression function could be modeled more flexibly rather than being restricted to a prespecified parametric form, potentially improving efficiency and broadening applicability. 
Third, the MSE selector is selection consistent under fixed alternatives with a strict risk gap, but no data-driven rule can consistently separate an $n^{-1/2}$-order local bias from sampling variation. Studying inference and smoother borrowing rules in this nonregular boundary regime remains important. \citet{hu2022semiparametric} establish semiparametric efficiency theory under the assumption that the internal and external populations have the same distribution. Developing analogous theory for the present setting, which allows the two populations to have different distributions, is an important direction for future research.

\section*{Acknowledgements}
The authors declare no conflict of interest. Reproducible simulation code and an analysis script for use with the approved real-data records are available at \url{https://github.com/KMorikawaISU/daisy}. The individual-level real data used in the application are not publicly available because of privacy restrictions and the potential identifiability of individuals; access may be granted on reasonable request, subject to approval by the data provider and relevant ethical and privacy constraints.
\par

\clearpage
\section*{Supplementary Material}
The supplementary material below contains the dual formulas and additional computation details used in the proposed entropy-balancing procedure, the regularity conditions and proofs for the asymptotic and efficiency results, the extension using externally reported regression coefficients, a bootstrap algorithm, and additional simulation results.

\setcounter{section}{0}
\setcounter{equation}{0}
\setcounter{figure}{0}
\setcounter{table}{0}
\setcounter{theorem}{0}
\setcounter{proposition}{0}
\setcounter{lemma}{0}
\setcounter{corollary}{0}
\renewcommand{\thesection}{S\arabic{section}}
\renewcommand{\theHsection}{S\arabic{section}}
\renewcommand{\theequation}{S\arabic{section}.\arabic{equation}}
\renewcommand{\theHequation}{S\arabic{section}.\arabic{equation}}
\renewcommand{\thefigure}{S\arabic{figure}}
\renewcommand{\theHfigure}{S.\arabic{figure}}
\renewcommand{\thetable}{S\arabic{table}}
\renewcommand{\theHtable}{S.\arabic{table}}
\renewcommand{\thetheorem}{S\arabic{section}.\arabic{theorem}}
\renewcommand{\theproposition}{S\arabic{section}.\arabic{proposition}}
\renewcommand{\thelemma}{S\arabic{section}.\arabic{lemma}}
\renewcommand{\thecorollary}{S\arabic{section}.\arabic{corollary}}
\makeatletter
\@addtoreset{theorem}{section}
\@addtoreset{proposition}{section}
\@addtoreset{lemma}{section}
\@addtoreset{corollary}{section}
\makeatother

\section{Technical Details}\label{supp:app:tech}
\vspace*{12pt}

\subsection{Regularity Conditions and Restated Results}

The following conditions supplement (C1)--(C3) and (C3)$'$ in the main paper. Write $\mu^*_{x\mid\mathrm{ex}}=\E_{\mathrm{ex}}(X)$ and $\eta^*_{\mathrm{ex}}=\E_{\mathrm{ex}}(Y)$, and let $\dot\rho_j$ denote the derivative of $\rho_j$. As in the main paper, EB denotes the estimator based on ordinary outcome regression, whereas EBW denotes its first-step-weighted-regression version.
\begin{itemize}
\item[(C4)] The probability limit $\mu^*_{x\mid\mathrm{ex}}$ belongs to the relative interior of the convex hull of the internal support of $X$. Let
\[
\mathcal H^*=\rho_1(\lambda_1^{*\top}\tilde X)\,\beta_{\mathrm{in}}^{*\top}\tilde X.
\]
Then
\[
\operatorname*{ess\,inf}\mathcal H^*<\eta^*_{\mathrm{ex}}<
\operatorname*{ess\,sup}\mathcal H^*,
\qquad \operatorname{Var}_{\mathrm{in}}(\mathcal H^*)>0.
\]
The population first-step, regression, and second-step equations have isolated solutions. Each solution has a compact local neighborhood on which the corresponding observation-level functions are continuous and dominated by an integrable envelope. The regression Jacobian is nonsingular, and the population balancing Hessians
\[
\E_{\mathrm{in}}\!\left[\dot\rho_1(\lambda_1^{*\top}\tilde X)\tilde X\tilde X^\top\right]
\quad\text{and}\quad
\E_{\mathrm{in}}\!\left[\dot\rho_2(\lambda_2^{*\top}\tilde{\mathcal H}^*)
\tilde{\mathcal H}^*\tilde{\mathcal H}^{*\top}\right]
\]
are positive definite, where $\tilde{\mathcal H}^*=(1,\mathcal H^*)^\top$.

\item[(C5)] Let
\[
\xi^*=(\beta_{\mathrm{in}}^{*\top},\lambda_1^{*\top},
\lambda_2^{*\top},\theta^*)^\top.
\]
On a compact neighborhood $\mathcal N$ of $\xi^*$, the observation-level estimating functions are continuously differentiable, and the functions and their first derivatives are bounded by integrable envelopes. The required $2+\delta$ moments are finite for some $\delta>0$. The external summaries satisfy a joint central limit theorem, and the population Jacobian $J(\xi^*)$ is nonsingular.
\end{itemize}

Conditions (C4)--(C5) require compactness only for a local neighborhood of the finite-dimensional parameter. They do not bound the realized weights. The integrable-envelope condition yields a uniform law of large numbers on $\mathcal N$, and the isolated-root condition localizes the estimator to that neighborhood with probability tending to one. Thus a global compact restriction on the weight values is unnecessary.

For reference, the results proved below are restated here.

\begin{proposition}\label{supp:prop:existence}
Under (C1), (C2), (C4), and (C3) or (C3)$'$, the two balancing problems have unique solutions with probability tending to one.
\end{proposition}

\begin{theorem}\label{supp:thm:consistency-s}
Under (C1), (C2), (C4), and (C3) or (C3)$'$, $\hat\theta_{\mathrm{EBW}}$ is consistent for $\theta^*$.
\end{theorem}

\begin{theorem}\label{supp:thm:clt-s}
Under (C1), (C2), (C4), (C5), and either (C3) or (C3)$'$,
\[
 n^{1/2}(\hat\theta_{\mathrm{EBW}}-\theta^*)
 \xrightarrow{d}N(0,\Sigma_{\mathrm{EBW}}).
\]
\end{theorem}

\begingroup
For the final mean squared error (MSE) selection theorem, let $\hat\theta_{\mathrm{I}}=\bar Y$ and let $\hat\theta_{\mathrm{P}}$ denote the proposed estimator after entropy-family selection. Write
\[
\theta_{\mathrm{P}}^\dagger=\operatorname*{plim}\hat\theta_{\mathrm{P}},
\qquad b_{\mathrm{P}}=\theta_{\mathrm{P}}^\dagger-\theta^*.
\]
For discussion of finite-sample and local misspecification, write
$b_{\mathrm{P},n}=\E(\hat\theta_{\mathrm{P}})-\theta^*$, where this expectation is over repeated sampling. Under the fixed alternatives considered in the theorem, $b_{\mathrm{P},n}\to b_{\mathrm{P}}$ whenever the root-$n$ sequence is uniformly integrable.
Let $\hat V_{\mathrm{I}}$, $\hat V_{\mathrm{P}}$, and $\hat V_{\mathrm{D}}$ estimate the variances of $\hat\theta_{\mathrm{I}}$, $\hat\theta_{\mathrm{P}}$, and $\hat\theta_{\mathrm{P}}-\hat\theta_{\mathrm{I}}$, respectively, with $\hat V_{\mathrm{D}}$ obtained from paired bootstrap replicates. For $c_n\to\infty$ and $c_n/\sqrt n\to0$, define
\begin{align}
\hat B_{\mathrm{P}}^2
={}&\left[(\hat\theta_{\mathrm{P}}-\hat\theta_{\mathrm{I}})^2-\hat V_{\mathrm{D}}\right]_+
\mathbf 1\left\{|\hat\theta_{\mathrm{P}}-\hat\theta_{\mathrm{I}}|>c_n\hat V_{\mathrm{D}}^{1/2}\right\},
\label{supp:eq:bias2-supp}
\\
\hat R_{\mathrm{I}}={}&\hat V_{\mathrm{I}},
\qquad
\hat R_{\mathrm{P}}=\hat V_{\mathrm{P}}+\hat B_{\mathrm{P}}^2.
\label{supp:eq:risk-supp}
\end{align}
The MSE selector chooses $\hat\theta_{\mathrm{P}}$ when $\hat R_{\mathrm{P}}\leq\hat R_{\mathrm{I}}$ and otherwise chooses $\hat\theta_{\mathrm{I}}$.

\begin{theorem}[Selection consistency]\label{supp:thm:mse-selection-s}
Suppose that $\hat\theta_{\mathrm{I}}-\theta^*=O_p(n^{-1/2})$ and $\hat\theta_{\mathrm{P}}-\theta_{\mathrm{P}}^\dagger=O_p(n^{-1/2})$, and that
\[
n\hat V_{\mathrm{I}}\xrightarrow{p}\Sigma_{\mathrm{I}},
\qquad
n\hat V_{\mathrm{P}}\xrightarrow{p}\Sigma_{\mathrm{P}},
\qquad
n\hat V_{\mathrm{D}}\xrightarrow{p}\Sigma_{\mathrm{D}}>0.
\]
If $b_{\mathrm{P}}=0$ and $\Sigma_{\mathrm{P}}\neq\Sigma_{\mathrm{I}}$, the rule selects the estimator with the smaller asymptotic variance with probability tending to one. If $b_{\mathrm{P}}\neq0$, it selects $\hat\theta_{\mathrm{I}}$ with probability tending to one.
\end{theorem}
\endgroup

The exact efficiency comparison involves the population least-squares slope
\begin{align}
\alpha
=\frac{\operatorname{Cov}_{\mathrm{in}}(Y,\mathcal H^*)}
       {\operatorname{Var}_{\mathrm{in}}(\mathcal H^*)}.
\label{supp:eq:alpha-supp}
\end{align}
This is a population quantity in the variance calculation, not an additional tuning parameter, and it is not used to construct either $\hat\theta_{\mathrm{EB}}$ or $\hat\theta_{\mathrm{EBW}}$. If $Y$ and $\mathcal H^*$ are square integrable under the internal distribution and $0<\operatorname{Var}_{\mathrm{in}}(\mathcal H^*)<\infty$, then Cauchy--Schwarz gives
\[
|\alpha|\leq
\left\{\frac{\operatorname{Var}_{\mathrm{in}}(Y)}{\operatorname{Var}_{\mathrm{in}}(\mathcal H^*)}\right\}^{1/2}<\infty.
\]
Thus $\alpha$ is finite under these conditions. Since it is a regression slope rather than a probability or correlation coefficient, it need not belong to $[0,1]$ and may exceed one. It is not uniformly bounded over sequences of data-generating processes along which $\operatorname{Var}_{\mathrm{in}}(\mathcal H^*)$ tends to zero. If $\operatorname{Var}_{\mathrm{in}}(\mathcal H^*)=0$, the second-step calibration direction is degenerate and $\alpha$ is not defined.

For the efficiency comparison, let $\mu^*_{x\mid\mathrm{in}}=\E_{\mathrm{in}}(X)$, $\Sigma^*=\operatorname{Var}_{\mathrm{in}}(X)$, and $r(x)=f_{\mathrm{ex}}(x)/f_{\mathrm{in}}(x)$, and define
\begin{align*}
D_1^2
&=(\mu^*_{x\mid\mathrm{ex}}-\mu^*_{x\mid\mathrm{in}})^\top
  \Sigma^{*-1}(\mu^*_{x\mid\mathrm{ex}}-\mu^*_{x\mid\mathrm{in}}),\\
D_2^2
&=\E_{\mathrm{in}}\{r(X)-1\}^2
  =\chi^2(f_{\mathrm{ex}}\|f_{\mathrm{in}}).
\end{align*}

\begin{theorem}\label{supp:thm:finite-kappa}
Suppose that (C1), (C2), (C3), and (C4)--(C6) hold and $n/n_1\to\kappa\in[0,\infty)$. Assume also that $Y$ and $\mathcal H^*$ are square integrable under the internal distribution and $0<\operatorname{Var}_{\mathrm{in}}(\mathcal H^*)<\infty$. If the regression coefficients are estimated by ordinary least squares, then
\begin{align}
\Sigma_{\mathrm{EB}}
&=\operatorname{Var}_{\mathrm{in}}(Y)-\sigma^2\alpha
\bigl[2-\alpha\{1+D_1^2+\kappa\}\bigr].
\label{supp:eq:finite-eb}
\end{align}
If, in addition, (C3)$'$ holds, $D_2^2<\infty$, and the regression coefficients are estimated by the first-step-weighted least-squares equation, then
\begin{align}
\Sigma_{\mathrm{EBW}}
&=\operatorname{Var}_{\mathrm{in}}(Y)-\sigma^2\alpha
\bigl[2-\alpha\{1+D_2^2+\kappa\}\bigr].
\label{supp:eq:finite-ebw}
\end{align}
Consequently, the corresponding estimator is at least as efficient as the internal sample mean if and only if
\[
\alpha\bigl[2-\alpha\{1+D_j^2+\kappa\}\bigr]\geq0,
\qquad j=1,2.
\]
\end{theorem}

\begingroup
\begin{remark}[No covariate heterogeneity]\label{supp:rem:no-heterogeneity-alpha}
If the internal and external covariate distributions are equal, then the population first-step weights satisfy $w^{*(1)}\equiv1$, so $\mathcal H^*$ equals the true conditional mean, or more generally the population least-squares projection with an intercept. Projection orthogonality gives
\[
\operatorname{Cov}_{\mathrm{in}}(Y,\mathcal H^*)=\operatorname{Var}_{\mathrm{in}}(\mathcal H^*),
\]
and hence $\alpha=1$. Thus equality of the covariate distributions is a sufficient, but not necessary, condition for $\alpha=1$. Substituting this value into the general criterion gives $D_1^2\leq1-\kappa$ and $D_2^2\leq1-\kappa$; when $\kappa=0$, these reduce to $D_1\leq1$ and $D_2^2\leq1$. Under exact homogeneity, both distances are themselves zero. Under covariate heterogeneity, $\alpha$ may be smaller than, equal to, or larger than one.
\end{remark}
\endgroup

\begin{remark}[Interpretation of the generalized criterion]
For $j=1$ (EB) or $j=2$ (EBW), the variance reduction in Theorem~\ref{supp:thm:finite-kappa} is
\[
\sigma^2\left[2\alpha-\alpha^2\{1+D_j^2+\kappa\}\right].
\]
For fixed $D_j$ and $\kappa$, the linear term represents the covariance reduction obtained from calibration, whereas the quadratic term represents the additional variation induced by population discrepancy and uncertainty in the external summaries. Hence, as $\alpha\downarrow0$, the non-inferiority condition becomes easier to satisfy but the gain itself vanishes. The expression is maximized at $\{1+D_j^2+\kappa\}^{-1}$ and is nonnegative precisely for $0\leq\alpha\leq2/\{1+D_j^2+\kappa\}$.
\end{remark}

\begin{example}[A case with $\alpha>1$]
Let $X$ be binary with $\Pr_{\mathrm{in}}(X=1)=p$ and $\Pr_{\mathrm{ex}}(X=1)=q$, where $p,q\in(0,1)$, and suppose $\E_{\mathrm{in}}(Y\mid X)=\E_{\mathrm{ex}}(Y\mid X)=X$. The density ratio satisfies $r(1)=q/p$, and therefore
\[
\mathcal H^*=r(X)X=\frac{q}{p}X.
\]
Since $\operatorname{Cov}_{\mathrm{in}}(Y,X)=\operatorname{Var}_{\mathrm{in}}(X)$ under the common conditional-mean model,
\[
\alpha
=\frac{\operatorname{Cov}_{\mathrm{in}}\{Y,(q/p)X\}}
       {\operatorname{Var}_{\mathrm{in}}\{(q/p)X\}}
=\frac{p}{q}.
\]
Thus $\alpha>1$ whenever the external prevalence $q$ is smaller than the internal prevalence $p$. In this example, $\alpha>1$ means that the second-step discrepancy is amplified. It can still produce a variance gain only when the discrepancy and $\kappa$ are sufficiently small to satisfy Theorem~\ref{supp:thm:finite-kappa}.
\end{example}

\begin{proposition}\label{supp:prop:selection}
Let the candidate set of first-step entropy families be finite, and suppose that (C4)--(C5) hold for each candidate. With $Q^*(j)=\E_{\mathrm{in}}\{G_2(w^{*(2)}_{[j]})\}$ and $\mathcal J_0=\arg\min_j Q^*(j)$, the entropy-family selector in Section~3.4 of the main paper satisfies $\Pr(\hat j\in\mathcal J_0)\to1$. If $\mathcal J_0$ is a singleton, the selector is consistent for that family.
\end{proposition}

\subsection{Existence and Uniqueness of the Balancing Weights}

The following relative-interior formulation is a standard convex-hull argument; see also Lemma~2 of \citet{zhao_entropy_2017}.

\begin{lemma}\label{supp:lem:convex-hull}
Let $V_1,\ldots,V_n$ be i.i.d. and let $t$ belong to the relative interior of $\operatorname{conv}\{\operatorname{supp}(V)\}$. If $\hat t\xrightarrow{p}t$, then
\[
\Pr\!\left\{\hat t\in\operatorname{ri}\operatorname{conv}(V_1,\ldots,V_n)\right\}\to1.
\]
\end{lemma}

\begin{proof}
Work in the affine hull of the support, whose dimension is denoted by $d$. Because $t$ is in the relative interior, there are $d+1$ affinely independent support points whose simplex contains $t$ in its relative interior. Small neighborhoods of these points can be chosen so that any selection of one point from each neighborhood still forms a simplex containing a fixed neighborhood of $t$. Each neighborhood has positive probability, so the sample intersects all of them with probability tending to one. Since $\hat t\to t$, the asserted event then follows.
\end{proof}

\begin{proof}[Proof of Proposition~\ref{supp:prop:existence}]
For the first step, apply Lemma~\ref{supp:lem:convex-hull} to $V=X$ and $\hat t=\hat\mu_{x\mid\mathrm{ex}}$. Hence the first-step target lies in the relative interior of the sample convex hull with probability tending to one. The normalization constraint and this interiority condition give strictly positive feasible weights. Because $G_1$ is strictly convex and its derivative ranges from $-\infty$ to $\infty$, standard convex duality gives a unique primal weight vector. The full-rank first-step Hessian in (C4) makes the dual criterion strictly convex in a neighborhood of its solution and therefore gives a unique multiplier $\hat\lambda_1$. The local uniform law and isolated-root conditions in (C4)--(C5) then give $\hat\lambda_1\xrightarrow{p}\lambda_1^*$; the corresponding regression conditions give $\hat\beta_{\mathrm{in}}\xrightarrow{p}\beta_{\mathrm{in}}^*$.

For the second step, the normalization constraint is separated from the scalar calibration constraint. Under (C3), the population first-step balance gives
$\E_{\mathrm{in}}(\mathcal H^*)=\beta_{\mathrm{in}}^{*\top}\tilde\mu^*_{x\mid\mathrm{ex}}=\eta^*_{\mathrm{ex}}$.
Under (C3)$'$, the first-step weight is the density ratio. By a change of measure and (C1),
\[
\E_{\mathrm{in}}\!\left[r(X)\tilde X\{Y-\beta^\top\tilde X\}\right]
=\E_{\mathrm{ex}}\!\left[\tilde X\{Y-\beta^\top\tilde X\}\right].
\]
Thus the population weighted regression equation based on the internal outcomes is the external-population normal equation without requiring equality of the full conditional distributions. The intercept in that equation gives
$\E_{\mathrm{in}}(\mathcal H^*)=\E_{\mathrm{ex}}(\beta_{\mathrm{in}}^{*\top}\tilde X)=\E_{\mathrm{ex}}(Y)=\eta^*_{\mathrm{ex}}$.
Condition (C4) strengthens this equality to strict interiority of the scalar target. Thus there are bounded sets of positive probability on which $\mathcal H^*$ is respectively below and above $\eta^*_{\mathrm{ex}}$ by a fixed margin. Consistency of $\hat\beta_{\mathrm{in}}$ and $\hat\lambda_1$, continuity of $\rho_1$, and consistency of $\hat\eta_{\mathrm{ex}}$ imply that, with probability tending to one, the observed values $\hat{\mathcal H}_i$ include values on both sides of $\hat\eta_{\mathrm{ex}}$. Hence the second-step target belongs to the interior of the interval generated by $\{\hat{\mathcal H}_i\}_{i=1}^n$. Strict convexity gives a unique positive second-step weight vector, and the full-rank second-step Hessian gives a unique multiplier $\hat\lambda_2$.
\end{proof}

\subsection{Double Robustness}

\begin{proof}[Proof of Theorem~\ref{supp:thm:consistency-s}]
First suppose that (C3) holds. \meanrev{Together with the conditional-mean equality in (C1),} the conditional mean-zero property of the regression error implies that the estimating equation weighted by the first-step weights has the same limit $\beta^*_{\mathrm{in}}=\beta^*_{\mathrm{ex}}$ as the correctly specified linear regression. The population first-step calibration equation then gives
\[
\E_{\mathrm{in}}(\mathcal H^*)
=\beta_{\mathrm{in}}^{*\top}\E_{\mathrm{in}}\!\left\{\rho_1(\lambda_1^{*\top}\tilde X)\tilde X\right\}
=\beta_{\mathrm{ex}}^{*\top}\tilde\mu^*_{x\mid\mathrm{ex}}
=\eta^*_{\mathrm{ex}}.
\]

Next suppose that (C3)$'$ holds, so that
$\rho_1(\lambda_1^{*\top}\tilde X)=f_{\mathrm{ex}}(X)/f_{\mathrm{in}}(X)$. A change of measure and (C1) give
\[
\E_{\mathrm{in}}\!\left[r(X)\tilde X\{Y-\beta^\top\tilde X\}\right]
=\E_{\mathrm{ex}}\!\left[\tilde X\{Y-\beta^\top\tilde X\}\right].
\]
Hence the weighted internal regression limit solves the external-population normal equation without requiring full conditional-distribution equality. Because an intercept is included,
\[
\E_{\mathrm{ex}}\{Y-\beta_{\mathrm{in}}^{*\top}\tilde X\}=0.
\]
It follows again that
\[
\E_{\mathrm{in}}(\mathcal H^*)
=\E_{\mathrm{ex}}(\beta_{\mathrm{in}}^{*\top}\tilde X)
=\E_{\mathrm{ex}}(Y)=\eta^*_{\mathrm{ex}}.
\]

In either case the population second-step constraints are satisfied by uniform weights. Hence $\rho_2(\lambda_2^{*\top}\tilde{\mathcal H}^*)=1$ almost surely, and the probability limit of the final estimating equation is
$\E_{\mathrm{in}}(\theta-Y)=0$. Therefore $\hat\theta_{\mathrm{EBW}}\xrightarrow{p}\theta^*$. \meanrev{This is the precise sense in which uniform second-step weights arise doubly robustly: either (C3) or (C3)$'$ yields the same population second-step balance. It is a consistency statement, not a doubly robust efficiency claim.}
\end{proof}

\subsection{Asymptotic Linear Representation}

Let
\[
\sqrt{n_1}
\begin{pmatrix}
\hat\mu_{x\mid\mathrm{ex}}-\mu^*_{x\mid\mathrm{ex}}\\
\hat\eta_{\mathrm{ex}}-\eta^*_{\mathrm{ex}}
\end{pmatrix}
\xrightarrow{d}
\begin{pmatrix}W_1\\W_2\end{pmatrix},
\]
independently of the internal sample. At the probability limit, define the internal contribution
\[
\mathcal I_i=
\begin{pmatrix}
\rho_1(\lambda_1^{*\top}\tilde X_i)\varphi(X_i,Y_i;\beta^*_{\mathrm{in}})\\
\rho_1(\lambda_1^{*\top}\tilde X_i)\tilde X_i-\tilde\mu^*_{x\mid\mathrm{ex}}\\
\tilde{\mathcal H}_i^*-\tilde\eta^*_{\mathrm{ex}}\\
\theta^*-Y_i
\end{pmatrix}.
\]
Let $J$ be the population Jacobian of the stacked equations and write the last row of $J^{-1}$ as
\[
(-\tau_0^\top,-\tau_1^\top,-\tau_2^\top,1).
\]

\begin{proof}[Proof of Theorem~\ref{supp:thm:clt-s}]
A first-order Taylor expansion of the stacked estimating equations around $\xi^*$ gives
\[
\sqrt n(\hat\xi-\xi^*)
=-J^{-1}\left[
\frac{1}{\sqrt n}\sum_{i=1}^n\mathcal I_i
-\sqrt\kappa
\begin{pmatrix}0\\W_1\\W_2\\0\end{pmatrix}
\right]+o_p(1).
\]
The last component is therefore
\begin{align*}
\sqrt n(\hat\theta_{\mathrm{EBW}}-\theta^*)
={}&\frac1{\sqrt n}\sum_{i=1}^n\Bigl[
Y_i-\theta^*
+\tau_2^\top(\tilde{\mathcal H}_i^*-\tilde\eta^*_{\mathrm{ex}})\\
&\qquad+\tau_1^\top\{\rho_1(\lambda_1^{*\top}\tilde X_i)\tilde X_i
-\tilde\mu^*_{x\mid\mathrm{ex}}\}\\
&\qquad+\tau_0^\top\rho_1(\lambda_1^{*\top}\tilde X_i)
\varphi(X_i,Y_i;\beta^*_{\mathrm{in}})
\Bigr]\\
&-\sqrt\kappa\,\tau_1^\top W_1\\
&-\sqrt\kappa\,\tau_2^\top W_2+o_p(1).
\end{align*}
Thus the internal influence function begins with $Y_i-\theta^*$, and each external-summary term enters exactly once. The multivariate central limit theorem and Slutsky's theorem yield the stated normal limit. The envelope and differentiability conditions in (C5) justify replacement of the sample Jacobian by $J$ uniformly on the local neighborhood.
\end{proof}

\subsection{Entropy-Family Selection}

\begin{proof}[Proof of Proposition~\ref{supp:prop:selection}]
For candidate $j$, let $w^{*(2)}_{[j]}$ be the probability limit of the second-step weight and define
\[
Q^*(j)=\E_{\mathrm{in}}\{G_2(w^{*(2)}_{[j]})\},
\qquad
\mathcal J_0=\arg\min_j Q^*(j).
\]
Because the candidate set is finite, (C5) and the uniform law of large numbers imply
\[
\max_j|\hat Q_n(j)-Q^*(j)|\xrightarrow{p}0.
\]
There is a positive gap between $\min_jQ^*(j)$ and every value outside $\mathcal J_0$, so $\Pr(\hat j\in\mathcal J_0)\to1$.

For any candidate whose limiting second-step weights are uniform, the normalization $\E_{\mathrm{in}}(w^{*(2)}_{[j]})=1$ and Jensen's inequality give
\[
Q^*(j)=\E_{\mathrm{in}}\{G_2(w^{*(2)}_{[j]})\}\geq G_2(1),
\]
with equality at $w^{*(2)}_{[j]}\equiv1$. Hence every such candidate belongs to $\mathcal J_0$. In particular, for EBW a correctly specified density-ratio candidate has uniform limiting second-step weights under (C1), by the double-robustness argument above. Several candidates may attain the same minimum, for example when the two populations coincide; this is why the result is stated for the minimizer set. If $\mathcal J_0$ is a singleton, conventional selection consistency follows.
\end{proof}

\begingroup
\subsection{Proof of the MSE-Selection Result}

\begin{proof}[Proof of Theorem~\ref{supp:thm:mse-selection-s}]
First suppose $b_{\mathrm{P}}=0$. Then
\[
\hat\theta_{\mathrm{P}}-\hat\theta_{\mathrm{I}}=O_p(n^{-1/2}),
\qquad
\hat V_{\mathrm{D}}^{1/2}=O_p(n^{-1/2}).
\]
Because $n\hat V_{\mathrm{D}}\to_p\Sigma_{\mathrm{D}}>0$, the standardized difference
\[
\frac{|\hat\theta_{\mathrm{P}}-\hat\theta_{\mathrm{I}}|}{\hat V_{\mathrm{D}}^{1/2}}
\]
is $O_p(1)$. Since $c_n\to\infty$, the indicator in \eqref{supp:eq:bias2-supp} is zero with probability tending to one. Therefore $\hat B_{\mathrm{P}}^2=0$ with probability tending to one, and
\[
n(\hat R_{\mathrm{P}}-\hat R_{\mathrm{I}})
\xrightarrow{p}\Sigma_{\mathrm{P}}-\Sigma_{\mathrm{I}}.
\]
A strict variance gap consequently determines the selected estimator with probability tending to one.

Now suppose $b_{\mathrm{P}}\neq0$. Root-$n$ consistency around the two probability limits gives
\[
\hat\theta_{\mathrm{P}}-\hat\theta_{\mathrm{I}}\xrightarrow{p}b_{\mathrm{P}}.
\]
Moreover,
\[
c_n\hat V_{\mathrm{D}}^{1/2}
=\frac{c_n}{\sqrt n}\{n\hat V_{\mathrm{D}}\}^{1/2}
=o_p(1),
\]
so the indicator in \eqref{supp:eq:bias2-supp} tends to one. Since $\hat V_{\mathrm{D}}=O_p(n^{-1})$,
\[
\hat B_{\mathrm{P}}^2\xrightarrow{p}b_{\mathrm{P}}^2.
\]
Thus $\hat R_{\mathrm{P}}\to_p b_{\mathrm{P}}^2>0$, while $\hat R_{\mathrm{I}}=O_p(n^{-1})\to_p0$, and the internal estimator is selected with probability tending to one.
\end{proof}

The stabilization is essential only near the unbiased case. Without it, the random quantity $(\hat\theta_{\mathrm{P}}-\hat\theta_{\mathrm{I}})^2-\hat V_{\mathrm{D}}$ remains of order $n^{-1}$ when $b_{\mathrm{P}}=0$ and can create spurious estimated bias. The theorem concerns fixed $b_{\mathrm{P}}=0$ or fixed $b_{\mathrm{P}}\neq0$. Under local misspecification $b_{\mathrm{P},n}=\delta/\sqrt n$, the bias and sampling error have the same order, so no selector based on one data set can achieve the same oracle-selection conclusion uniformly.
\endgroup

\subsection{Efficiency Calculation and the General \texorpdfstring{$(\alpha,\kappa)$}{(alpha,kappa)} Criterion}

Write
\[
Y=\beta_{\mathrm{in}}^{*\top}\tilde X+\varepsilon,
\qquad \E_{\mathrm{in}}(\varepsilon\mid X)=0,
\qquad \E_{\mathrm{in}}(\varepsilon^2\mid X)=\sigma^2.
\]
The second-step calibration equation is locally a one-dimensional calibration on $\mathcal H^*$. Linearizing it around the uniform second-step weights gives
\begin{align}
\hat\theta
=\bar Y+\alpha
\left\{\hat\eta_{\mathrm{ex}}
-n^{-1}\sum_{i=1}^n\hat{\mathcal H}_i\right\}
+o_p(n^{-1/2}),
\label{supp:eq:second-step-linearization}
\end{align}
where $\alpha$ is defined in \eqref{supp:eq:alpha-supp}. To see this, the nonconstant component of $-\tau_2$ in the asymptotic linear representation equals
\[
\frac{\operatorname{Cov}_{\mathrm{in}}(Y,\mathcal H^*)}
{\operatorname{Var}_{\mathrm{in}}(\mathcal H^*)}=\alpha.
\]
The common factor $\dot\rho_2\{\rho_2^{-1}(1)\}$ cancels from the relevant Jacobian ratio, so the coefficient does not depend on the particular strictly convex second-step entropy.

For EB, let
\[
M_{\mathrm{in}}=\E_{\mathrm{in}}(\tilde X\tilde X^\top),
\qquad
c_1(X)=\tilde\mu_{x\mid\mathrm{ex}}^{*\top}
M_{\mathrm{in}}^{-1}\tilde X.
\]
The first-step calibration equation implies the exact identity
\[
n^{-1}\sum_{i=1}^n\hat{\mathcal H}_i
=\hat\beta_{\mathrm{in}}^\top\tilde\mu_{x\mid\mathrm{ex}}.
\]
Combining the ordinary least-squares expansion with the external-summary central limit theorem yields
\begin{align}
\sqrt n(\hat\theta_{\mathrm{EB}}-\theta^*)
={}&\frac1{\sqrt n}\sum_{i=1}^n
\{Y_i-\theta^*-\alpha c_1(X_i)\varepsilon_i\}
+\alpha\sqrt\kappa\,W_\varepsilon+o_p(1),
\label{supp:eq:eb-expansion}
\end{align}
where
\[
W_\varepsilon
=\lim\sqrt{n_1}\left[
(\hat\eta_{\mathrm{ex}}-\eta_{\mathrm{ex}}^*)
-\beta_{\mathrm{in}}^{*\top}
(\tilde\mu_{x\mid\mathrm{ex}}-\tilde\mu_{x\mid\mathrm{ex}}^*)
\right].
\]
\meanrev{Conditions (C1) and (C3) give the external residual mean zero, while the common source-specific variance in (C6) gives $\operatorname{Var}(W_\varepsilon)=\sigma^2$. If $\kappa=0$, this external variance condition is not used.} Since an intercept is included,
\[
\E_{\mathrm{in}}\{c_1(X)\}=1,
\qquad
\E_{\mathrm{in}}\{c_1(X)^2\}=1+D_1^2.
\]
Moreover,
\[
\operatorname{Cov}_{\mathrm{in}}\{Y-\theta^*,c_1(X)\varepsilon\}=\sigma^2,
\qquad
\operatorname{Var}_{\mathrm{in}}\{c_1(X)\varepsilon\}=\sigma^2(1+D_1^2).
\]
Independence of the internal and external samples then gives
\[
\Sigma_{\mathrm{EB}}
=\operatorname{Var}_{\mathrm{in}}(Y)-\sigma^2\alpha
\bigl[2-\alpha\{1+D_1^2+\kappa\}\bigr].
\]
This calculation uses the correct linear outcome model and exact first-step mean balance, but not a correctly specified density-ratio model.

For EBW, under (C3)$'$, the weighted least-squares expansion is
\[
\sqrt n(\hat\beta_{\mathrm{in}}-\beta^*_{\mathrm{in}})
=
\left[\E_{\mathrm{in}}\{r(X)\tilde X\tilde X^\top\}\right]^{-1}
\frac1{\sqrt n}\sum_{i=1}^n
r(X_i)\tilde X_i\varepsilon_i+o_p(1).
\]
Estimating $\lambda_1$ has no additional first-order effect because the derivative of the weighted regression equation with respect to $\lambda_1$ has conditional expectation zero under (C3). Since
\[
\tilde\mu_{x\mid\mathrm{ex}}^{*\top}
\left[\E_{\mathrm{in}}\{r(X)\tilde X\tilde X^\top\}\right]^{-1}
=e_1^\top,
\]
where $e_1=(1,0,\ldots,0)^\top$, we obtain
\begin{align}
\sqrt n(\hat\theta_{\mathrm{EBW}}-\theta^*)
={}&\frac1{\sqrt n}\sum_{i=1}^n
\{Y_i-\theta^*-\alpha r(X_i)\varepsilon_i\}
+\alpha\sqrt\kappa\,W_\varepsilon+o_p(1).
\label{supp:eq:ebw-expansion}
\end{align}
Because $\E_{\mathrm{in}}\{r(X)\}=1$ and
\[
\E_{\mathrm{in}}\{r(X)^2\}=1+D_2^2,
\]
we have
\[
\operatorname{Cov}_{\mathrm{in}}\{Y-\theta^*,r(X)\varepsilon\}=\sigma^2,
\qquad
\operatorname{Var}_{\mathrm{in}}\{r(X)\varepsilon\}=\sigma^2(1+D_2^2).
\]
Equation~\eqref{supp:eq:ebw-expansion} therefore gives
\[
\Sigma_{\mathrm{EBW}}
=\operatorname{Var}_{\mathrm{in}}(Y)-\sigma^2\alpha
\bigl[2-\alpha\{1+D_2^2+\kappa\}\bigr],
\]
which proves Theorem~\ref{supp:thm:finite-kappa}.

For either estimator, if $\alpha>0$, the exact condition is
\[
D_j^2\leq\frac{2}{\alpha}-1-\kappa,
\qquad j=1,2.
\]
If $\alpha=0$, the first-order variances are equal. If $\alpha<0$, the variance difference is negative because $1+D_j^2+\kappa>0$. When the covariate distributions coincide, Remark~\ref{supp:rem:no-heterogeneity-alpha} supplies $\alpha=1$.

Write $\mathbb P_n h=n^{-1}\sum_{i=1}^n h_i$ and $\bar{\mathcal H}=n^{-1}\sum_{i=1}^n\hat{\mathcal H}_i$. A plug-in estimator of $\alpha$ can be written as
\begin{align}
\hat\alpha
=\frac{\mathbb P_n\{(Y-\bar Y)(\hat{\mathcal H}-\bar{\mathcal H})\}}
{\mathbb P_n\{(\hat{\mathcal H}-\bar{\mathcal H})^2\}}.
\label{supp:eq:alpha-hat-supp}
\end{align}
Under (C4)--(C5),
$\mathbb P_n(\hat{\mathcal H}-\mathcal H^*)^2=o_p(1)$.
The numerator and denominator in \eqref{supp:eq:alpha-hat-supp} therefore converge to
$\operatorname{Cov}_{\mathrm{in}}(Y,\mathcal H^*)$ and
$\operatorname{Var}_{\mathrm{in}}(\mathcal H^*)$, respectively. Since the latter is positive, the continuous mapping theorem gives $\hat\alpha\xrightarrow{p}\alpha$.

\begingroup
For descriptive use, the general efficiency criterion can be evaluated by replacing $\alpha$, $D_j$, and $\kappa$ with $\hat\alpha$, $\hat D_j$, and $n/n_1$. This plug-in calculation is reported as an efficiency diagnostic rather than used for the final borrowing decision, which is made by the MSE selector. Since $\alpha$ is not bounded above by one, $\hat\alpha$ is not truncated at one. For EBW,
\endgroup
\[
\hat D_2^2
=n^{-1}\sum_{i=1}^n\{\hat w_i^{(1)}-1\}^2
=n^{-1}\sum_{i=1}^n\{\hat w_i^{(1)}\}^2-1
\]
because the first-step weights are normalized to have sample mean one. If the sample variance of $\hat{\mathcal H}$ is numerically negligible, the rule returns the internal sample mean rather than using an unstable slope estimate.

\begin{lemma}\label{supp:lem:distance}
The squared Mahalanobis distance satisfies $D_1^2\leq D_2^2$.
\end{lemma}

\begin{proof}
By the definitions of $r$ and $\mu_{x\mid\mathrm{in}}^*$,
\[
\mu_{x\mid\mathrm{ex}}^*-\mu_{x\mid\mathrm{in}}^*
=\E_{\mathrm{in}}\bigl[(X-\mu_{x\mid\mathrm{in}}^*)\{r(X)-1\}\bigr].
\]
For any vector $v$, Cauchy--Schwarz gives
\[
\{v^\top(\mu_{x\mid\mathrm{ex}}^*-\mu_{x\mid\mathrm{in}}^*)\}^2
\leq
(v^\top\Sigma^*v)\,\E_{\mathrm{in}}\{r(X)-1\}^2.
\]
Maximizing the left-hand ratio over $v$ proves the result.
\end{proof}

When the internal and external covariate distributions are Gaussian with a common covariance,
$D_2^2=\exp(D_1^2)-1$. For $\alpha>0$, the general EBW rule is therefore
\[
D_1\leq\left\{\log\left(\frac{2}{\alpha}-\kappa\right)\right\}^{1/2},
\]
provided $2/\alpha-\kappa\geq1$. Under no covariate heterogeneity, $\alpha=1$ and both distances equal zero. Algebraically, the $\alpha=1$, $\kappa=0$ Gaussian criterion has threshold $D_1\leq\sqrt{\log2}$.

\section{Computation Details}\label{supp:sec:S2}
\setcounter{equation}{0}
\label{supp:sec:selection}
\label{supp:sec:3.3}

This section records the entropy families and dual formulas used in computation. We keep the notation from the main paper: $g=G'$ denotes the derivative of the entropy function, $\rho=g^{-1}$ is the inverse link generating the balancing weights, and $F$ is the Fenchel conjugate.

\subsection{Entropy Families}
Besides the canonical KL entropy $G(w)=w\log w-w$, the numerical work considers the Lambert--$W$ (LW), quadratic log-sum (QLS), and tempered-softplus (TS) families below, with tuning parameters $a,r,c>0$. Their inverse links are explicit, which makes the dual optimization straightforward.
\begin{table}[htbp]
\centering
\scriptsize
\setlength{\tabcolsep}{3pt}
\renewcommand{\arraystretch}{1.12}
\resizebox{\textwidth}{!}{%
\begin{tabular}{lllll}
\toprule
Entropy & $G(w)$ & $g(w)$ & $\rho(x)$ & $F(x)$ \\
\midrule
$G_L$
& $w\log w-w+aw^2/2$
& $\log w+aw$
& $a^{-1}W_0\{a\exp(x)\}$
& $\{W_0(a\exp x)^2+2W_0(a\exp x)\}/(2a)$ \\
$G_Q$
& $w\log w-w+\{(1+rw)\log(1+rw)-(1+rw)\}/r$
& $\log w+\log(1+rw)$
& $\{\sqrt{1+4r\exp(x)}-1\}/(2r)$
& $[\sqrt{1+4r\exp(x)}-\log\{(1+\sqrt{1+4r\exp(x)})/2\}]/r$ \\
$G_S$
& $w^2/2+c^{-2}\operatorname{Li}_2\{\exp(-cw)\}$
& $c^{-1}\log\{\exp(cw)-1\}$
& $c^{-1}\log\{1+\exp(cx)\}$
& $-c^{-2}\operatorname{Li}_2\{-\exp(cx)\}-\pi^2/(6c^2)$ \\
\bottomrule
\end{tabular}%
}
\caption{Entropy functions, derivatives, inverse links, and Fenchel conjugates used in the candidate set. $W_0$ is the principal Lambert--$W$ branch and $\operatorname{Li}_2$ is the dilogarithm.}
\label{supp:tab:entropy-supp}
\end{table}
For the Lambert--$W$ family, $G_L''(w)=1/w+a>0$ and the inverse of $g_L(w)=\log w+aw$ is $\rho_L(x)=a^{-1}W_0\{a\exp(x)\}$. As $a\downarrow0$, this family approaches the canonical KL entropy and its exponential inverse link. The other two families provide alternative curvature and tail tempering. In the Gaussian case, balancing $(1,x,x^2)$ with the KL inverse link represents the exponential-quadratic density ratio exactly.

\subsection{Dual Formulation}
For completeness, we restate the dual derivation for the first-step balancing problem. The primal problem is
\begin{align*}
\max_{w^{(1)}}\ -n^{-1}\sum_{i=1}^n G_1(w^{(1)}_i)
\end{align*}
subject to
\begin{align}
n^{-1}\sum_{i=1}^n w^{(1)}_i = 1, \quad 
n^{-1}\sum_{i=1}^n w^{(1)}_i X_i = \hat{\mu}_{x\mid\mathrm{ex}}.
\label{supp:EB1}
\end{align}
Its dual problem is
\begin{align}
\min_{\lambda \in \mathbb{R}^d}\ \Biggl\{ n^{-1}\sum_{i=1}^n F\!\bigl(\lambda^\top \tilde{X}_i\bigr)-\lambda^\top \tilde{\mu}_{x\mid\mathrm{ex}} \Biggr\},
\label{supp:dual}
\end{align}
where $d=1+\dim(X)$. Since $dF(u)/du=\rho(u)$, the first-order conditions are
\begin{align}
n^{-1} \sum_{i=1}^n \rho(\lambda^\top \tilde{X}_i) \tilde{X}_i = \tilde{\mu}_{x\mid\mathrm{ex}},
\label{supp:g.const}
\end{align}
so the weights are $\hat w_i=\rho(\hat\lambda^\top\tilde X_i)$. The two balancing equations used in the main paper are therefore
\begin{align}
\ell_1(\hat{\lambda}_1) &= n^{-1} \sum_{i=1}^n \rho_1(\hat{\lambda}_1^\top \tilde{X}_i) \tilde{X}_i - \tilde{\mu}_{x\mid\mathrm{ex}} = 0, \label{supp:est_eq1} \\
\ell_2(\hat{\beta}_{\mathrm{in}}, \hat{\lambda}_1, \hat{\lambda}_2)
&= n^{-1} \sum_{i=1}^n \rho_2(\hat{\lambda}_2^\top \tilde{\mathcal{H}}_i) \tilde{\mathcal{H}}_i - \tilde{\eta}_{\mathrm{ex}} = 0, \label{supp:est_eq2}
\end{align}
with $\tilde{\mathcal{H}}_i = (1, \hat{w}^{(1)}_i(\hat{\lambda}_1) \hat{\eta}_{i\mid\mathrm{in}})^\top$, and the estimating equation for the final parameter is
\begin{align}
\ell_3(\hat{\beta}_{\mathrm{in}}, \hat{\lambda}_1,\hat{\lambda}_2, \hat{\theta}_{\mathrm{EBW}})=n^{-1}\sum_{i=1}^n \rho_2(\hat{\lambda}_2^\top\tilde{\mathcal{H}}_i)( \hat{\theta}_{\mathrm{EBW}} - Y_i )=0.
\label{supp:est_eq3}
\end{align}
The model-selection criterion for the first-step entropy family is
\begin{align}
 \hat{Q}_n(j)= n^{-1}\sum_{i=1}^n G_2\!\bigl(\hat w^{(2)}_{i[j]}\bigr), \label{supp:select}
\end{align}
where the selected family minimizes $\hat Q_n(j)$ over the candidate set.

\section{Extension beyond the Linear Baseline with Externally Reported Regression Information}\label{supp:sec:3.5}
\setcounter{equation}{0}

The mean-only baseline in the main paper uses a linear working regression because
\[
\E_{\mathrm{ex}}(\beta^\top\tilde X)
=\beta^\top\tilde\mu_{x\mid\mathrm{ex}},
\]
so the external average prediction is identified from the reported covariate means. For a nonlinear working mean $m\{\eta(X;\beta)\}$, where $m$ is a known link function, regression coefficients and $\mu_{x\mid\mathrm{ex}}$ alone are generally insufficient, since
\[
\E_{\mathrm{ex}}[m\{\eta(X;\beta)\}]
\neq m\{\eta(\E_{\mathrm{ex}}X;\beta)\}
\]
in general.

The two-step construction nevertheless extends directly when the external source reports enough information to identify the external average prediction. Two examples are:
\begin{enumerate}
\item an externally computed average fitted response
\[
\hat\eta_{\mathrm{ex}}
=n_1^{-1}\sum_{i=n+1}^{n+n_1}m\{\eta(X_i;\hat\beta_{\mathrm{ex}})\};
\]
\item external means of a basis vector $h(X)$ when the working mean is linear in that basis, so that
$\hat\eta_{\mathrm{ex}}=\hat\beta_{\mathrm{ex}}^\top(1,\overline{h}_{\mathrm{ex}}^\top)^\top$.
\end{enumerate}
In either case, replace the external outcome mean in the second-step constraint by the identified $\hat\eta_{\mathrm{ex}}$ and otherwise retain the same equations. This extension requires a working mean and sufficient external summaries, but not a full parametric specification of the conditional distribution of $Y\mid X$.

\section{Bootstrap Variance Estimation and MSE Selection}\label{supp:app:bootstrap}
\setcounter{equation}{0}

\begingroup
The final MSE criterion requires the variance of the proposed estimator and the variance of its difference from the internal sample mean. These quantities must be estimated jointly. In every bootstrap replicate, the same resampled internal observations are therefore used to compute the pair
\[
(\hat\theta_{\mathrm{I}}^*,\hat\theta_{\mathrm{P}}^*).
\]
The estimators are
\begin{align*}
\hat V_{\mathrm{P}}
&=\operatorname{Var}_*(\hat\theta_{\mathrm{P}}^*),\\
\hat V_{\mathrm{D}}
&=\operatorname{Var}_*(\hat\theta_{\mathrm{P}}^*-\hat\theta_{\mathrm{I}}^*),\\
\hat V_{\mathrm{I}}
&=s_Y^2/n,
\qquad
s_Y^2=(n-1)^{-1}\sum_{i=1}^n(Y_i-\bar Y)^2.
\end{align*}
Using separate bootstrap samples for the two estimators would lose their covariance and would not consistently estimate $\hat V_{\mathrm{D}}$.

When external individual-level data are available to the analyst, as in the simulation and the registry analysis, we independently bootstrap the internal and external samples and recompute the external summaries in each replicate. Only the recomputed summaries enter the estimator. This directly propagates uncertainty in both sources and avoids a separate nested bootstrap for estimating the covariance of the external summaries. For computational efficiency, the first-step entropy family selected from the original sample is held fixed during this variance bootstrap; only the final MSE comparison is performed after the bootstrap variances are obtained.

When only external summaries are available, let
\[
\hat t_{\mathrm{ex}}=(\hat\mu_{x\mid\mathrm{ex}}^\top,\hat\eta_{\mathrm{ex}})^\top
\]
and suppose a consistent estimator $\hat\Omega_{\mathrm{ex}}$ of the joint asymptotic covariance of $\sqrt{n_1}(\hat t_{\mathrm{ex}}-t_{\mathrm{ex}}^*)$ is available. Within each internal bootstrap replicate draw
\[
t_{\mathrm{ex}}^*
=\hat t_{\mathrm{ex}}+n_1^{-1/2}\hat\Omega_{\mathrm{ex}}^{1/2}Z,
\qquad Z\sim N(0,I),
\]
and recompute $\hat\theta_{\mathrm{P}}^*$. Marginal standard errors alone do not determine $\hat\Omega_{\mathrm{ex}}$ unless an additional independence approximation is imposed.

For confidence intervals after hard selection, the entropy-family selection and the MSE decision would need to be repeated within each bootstrap replicate. The numerical study focuses on point-estimation risk and uses the paired bootstrap above to construct the MSE decision. The intervals displayed in the application are component-wise bootstrap intervals for the estimator selected in the observed data; they are not claimed to have postselection coverage.
\endgroup

\section{Additional Numerical Results}
\setcounter{equation}{0}

\begingroup
\subsection{Simulation Design Indexed by Mahalanobis Distance}

The simulation uses 1,000 Monte Carlo repetitions, $n=200$, $n_1=2{,}000$, $X_1\sim N(0,1)$ and $X_2\sim\operatorname{Bernoulli}(0.5)$ internally, with calibration vector $B(X)=(X_1,X_1^2,X_2)^\top$. Since
\[
\operatorname{Var}_{\mathrm{in}}\{B(X)\}=\operatorname{diag}(1,2,1/4),
\]
the target Mahalanobis distance is determined by the first two moments of $X_1$ when the distribution of $X_2$ is unchanged. This distance indexes the designed degree of population heterogeneity and is not used in the final MSE selection. For $D\in\{0,0.25,0.5,1,1.5\}$, the external designs are:
\begin{align*}
&\text{mean shift: }X_1\sim N(\mu_D,1),\quad
\mu_D=\{-1+\sqrt{1+2D^2}\}^{1/2},\\
&\text{variance shift: }X_1\sim N(0,\sigma_D^2),\quad
\sigma_D^2=1+\sqrt2D,\\
&\text{Gamma shift: }X_1=\mu_D+(G-4)/2,\quad G\sim\operatorname{Gamma}(4,1).
\end{align*}
At $D=0$, the standardized Gamma distribution matches the first two normal moments but differs in shape. Outcomes satisfy
\[
Y\mid X_1,X_2\sim N(0.5X_1-X_2+0.5aX_1^2,1),\qquad a\in\{0,1\}.
\]
The target internal means are $-0.5$ for $a=0$ and $0$ for $a=1$. Here SM denotes the internal sample mean, EB and EBW denote the proposed estimators based on ordinary and first-step-weighted outcome regressions, respectively, S1EF denotes the estimator obtained after selecting the first-step entropy family by the criterion in Section~3.4 of the main paper, and MSE-Select denotes the final comparison between the corresponding S1EF estimator and SM using $c_n=\sqrt{\log n}$. The paired bootstrap independently resamples the internal and external individual-level samples.

\begin{figure}[htbp]
\centering
\includegraphics[width=\linewidth]{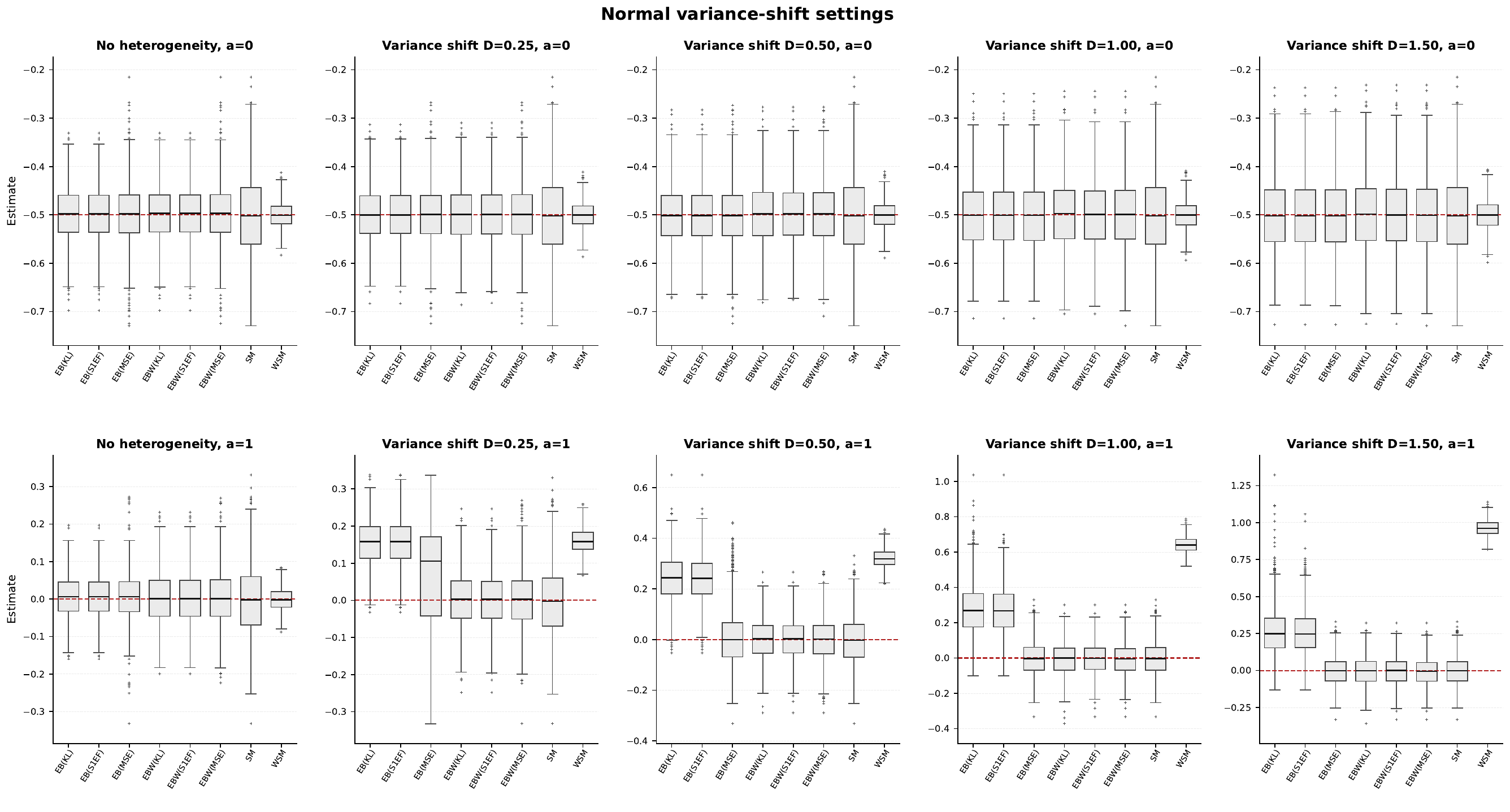}
\caption{Monte Carlo distributions under no heterogeneity and normal variance shifts. Top row: $a=0$; bottom row: $a=1$. Dashed lines mark the true internal-population mean.}
\label{supp:fig:S-variance-mse}
\end{figure}

\begin{figure}[htbp]
\centering
\includegraphics[width=\linewidth]{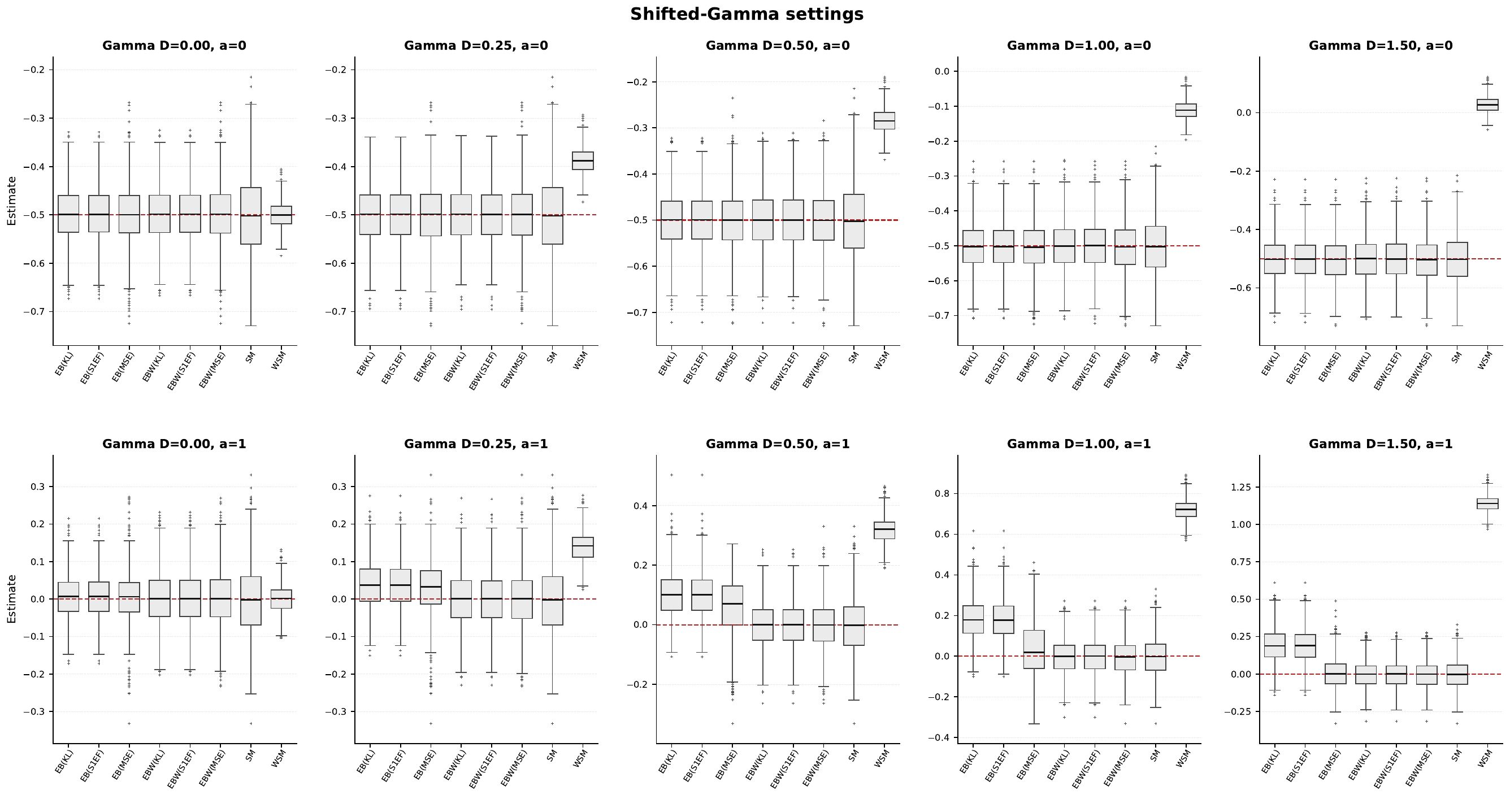}
\caption{Monte Carlo distributions under shifted-Gamma external covariates. Top row: $a=0$; bottom row: $a=1$. Dashed lines mark the true internal-population mean.}
\label{supp:fig:S-gamma-mse}
\end{figure}

\IfFileExists{FigureS_MSE_borrow_frequency.pdf}{%
\begin{figure}[htbp]
\centering
\includegraphics[width=\linewidth]{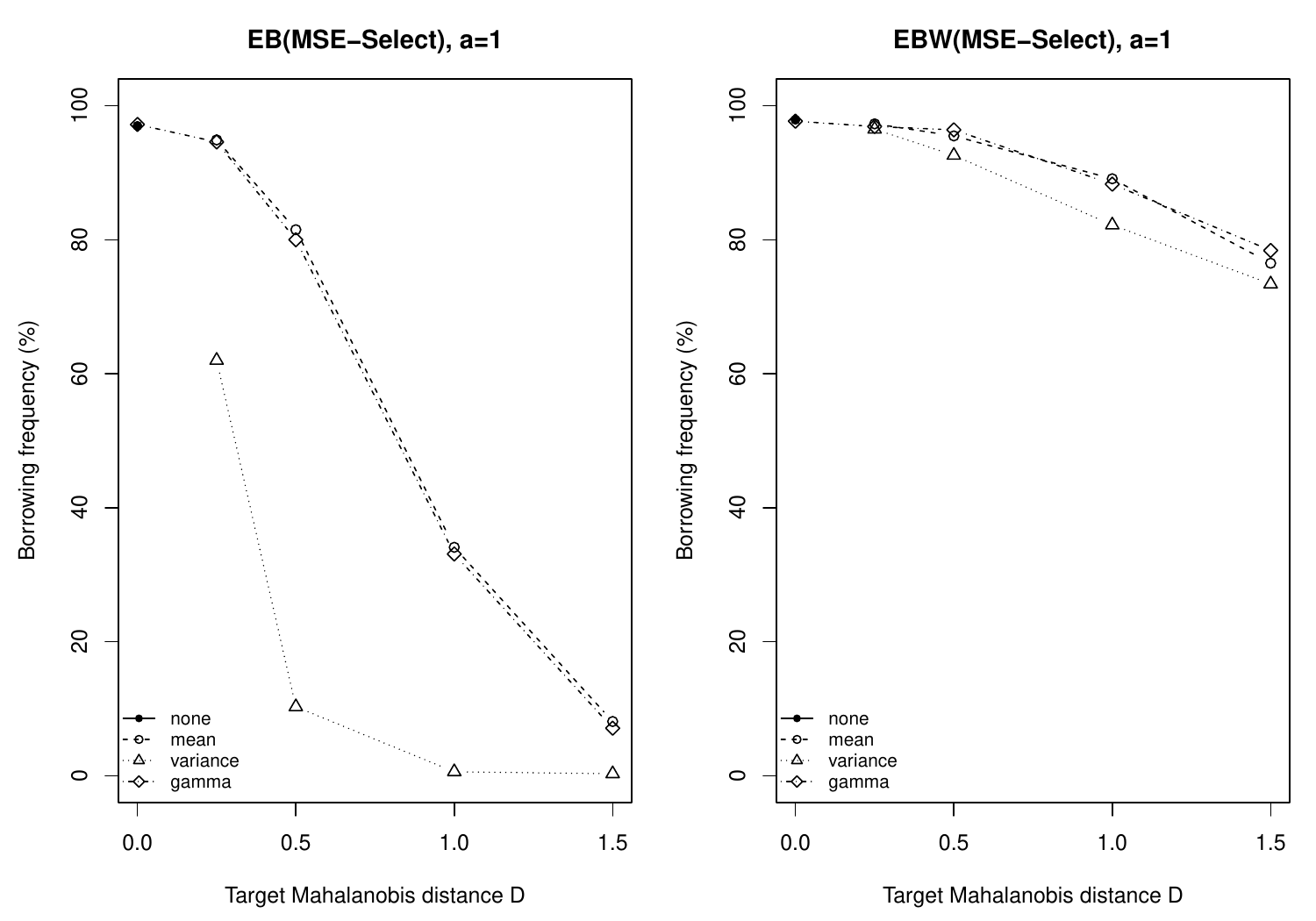}
\caption{Borrowing frequency of the MSE selectors when the outcome model is misspecified ($a=1$). EBW retains borrowing more often than EB because its weighted-regression construction remains stable under the simulated covariate shifts.}
\label{supp:fig:mse-borrow}
\end{figure}
}{}

\begingroup
\scriptsize
\setlength{\tabcolsep}{2.7pt}
\renewcommand{\arraystretch}{1.08}
\begin{longtable}{clrrrrr}
\caption{Monte Carlo performance of the final MSE selectors. EB and EBW use ordinary and first-step-weighted outcome regressions, respectively; MSE ratios are relative to the internal sample mean (SM).}\label{supp:tab:mse-simulation}\\
\toprule
$a$ & Shift & $D$ & \shortstack{EB\\MSE ratio} & \shortstack{EB\\borrow (\%)} & \shortstack{EBW\\MSE ratio} & \shortstack{EBW\\borrow (\%)} \\ \midrule
\endfirsthead
\toprule
$a$ & Shift & $D$ & \shortstack{EB\\MSE ratio} & \shortstack{EB\\borrow (\%)} & \shortstack{EBW\\MSE ratio} & \shortstack{EBW\\borrow (\%)} \\ \midrule
\endhead
0 & None & 0.00 & 0.518 & 97.9 & 0.507 & 98.2 \\
0 & Mean & 0.25 & 0.561 & 97.4 & 0.550 & 97.6 \\
0 & Mean & 0.50 & 0.585 & 97.9 & 0.607 & 97.6 \\
0 & Mean & 1.00 & 0.713 & 97.6 & 0.786 & 94.1 \\
0 & Mean & 1.50 & 0.813 & 91.9 & 0.890 & 83.5 \\
0 & Variance & 0.25 & 0.514 & 98.5 & 0.532 & 98.7 \\
0 & Variance & 0.50 & 0.585 & 99.1 & 0.640 & 98.3 \\
0 & Variance & 1.00 & 0.745 & 96.5 & 0.817 & 94.2 \\
0 & Variance & 1.50 & 0.855 & 89.5 & 0.927 & 82.5 \\
0 & Gamma & 0.00 & 0.502 & 97.9 & 0.484 & 98.4 \\
0 & Gamma & 0.25 & 0.543 & 97.6 & 0.546 & 97.5 \\
0 & Gamma & 0.50 & 0.586 & 98.0 & 0.578 & 97.9 \\
0 & Gamma & 1.00 & 0.715 & 96.6 & 0.784 & 94.0 \\
0 & Gamma & 1.50 & 0.821 & 91.7 & 0.904 & 83.3 \\
1 & None & 0.00 & 0.450 & 97.0 & 0.581 & 97.9 \\
1 & Mean & 0.25 & 0.645 & 94.9 & 0.632 & 97.3 \\
1 & Mean & 0.50 & 1.497 & 81.5 & 0.702 & 95.5 \\
1 & Mean & 1.00 & 1.954 & 34.1 & 0.851 & 89.1 \\
1 & Mean & 1.50 & 1.236 & 8.1 & 0.924 & 76.5 \\
1 & Variance & 0.25 & 2.289 & 62.0 & 0.652 & 96.5 \\
1 & Variance & 0.50 & 1.597 & 10.3 & 0.744 & 92.6 \\
1 & Variance & 1.00 & 1.019 & 0.6 & 0.877 & 82.2 \\
1 & Variance & 1.50 & 1.005 & 0.3 & 0.941 & 73.4 \\
1 & Gamma & 0.00 & 0.472 & 97.2 & 0.608 & 97.7 \\
1 & Gamma & 0.25 & 0.656 & 94.6 & 0.641 & 96.9 \\
1 & Gamma & 0.50 & 1.388 & 80.0 & 0.696 & 96.4 \\
1 & Gamma & 1.00 & 1.959 & 33.1 & 0.849 & 88.3 \\
1 & Gamma & 1.50 & 1.218 & 7.1 & 0.930 & 78.4 \\
\bottomrule
\end{longtable}
\endgroup

\subsection{Focused Selection-Consistency Experiment}

To assess the selection-consistency theorem in Section~5.1 of the main paper directly, we used 1,000 Monte Carlo repetitions for each $n\in\{200,500,1000\}$ with $n_1=10n$. In the correctly specified experiment, the external distribution was a normal mean shift with $D=0.5$ and $Y\mid X_1,X_2\sim N(0.5X_1-X_2,1)$. In the dual-misspecification experiment, the external covariate had a standardized Gamma shape shift with $D=0$, while $Y\mid X_1,X_2\sim N(0.5X_1-X_2+0.25X_1^3,1)$ and the working regression omitted the cubic term. Thus the moment-based Mahalanobis distance was zero although the selected EBW estimator had a fixed bias. Figure~\ref{supp:fig:S-robustness} compares SM, EBW(S1EF), the MSE selectors based on $\sqrt{\log n}$ and $\{\log\log(n+e^e)\}^{1/2}$, and an oracle selector.

\begin{figure}[htbp]
\centering
\includegraphics[width=\linewidth]{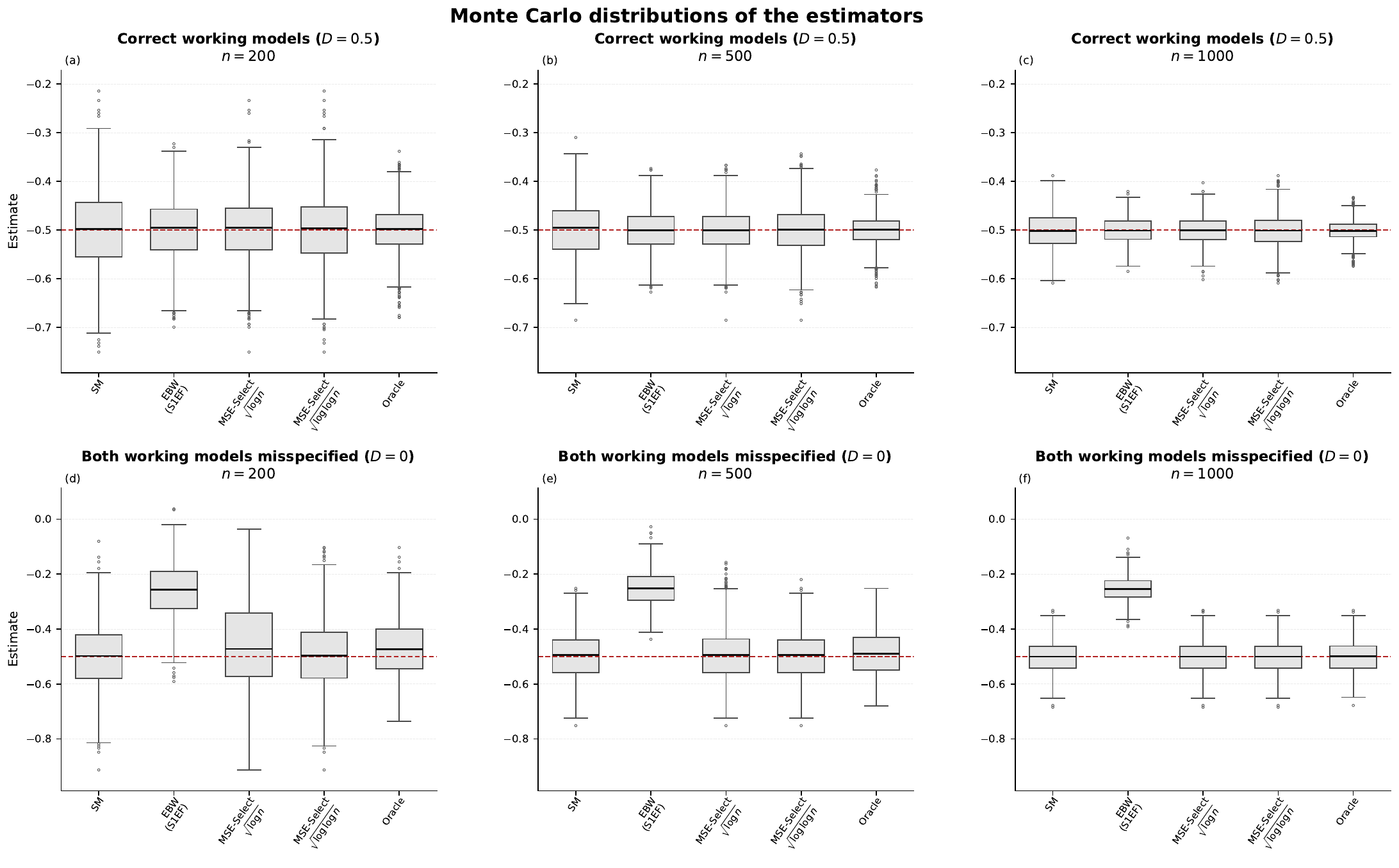}
\caption{Selection-consistency experiment. Under correct working models (top row), the MSE selector increasingly uses EBW. Under dual misspecification (bottom row), it increasingly reverts to SM even though the Mahalanobis distance equals zero.}
\label{supp:fig:S-robustness}
\end{figure}

For the primary choice $c_n=\sqrt{\log n}$, the borrowing probabilities under correct models were $97.8\%$, $99.0\%$, and $99.2\%$, with MSE ratios relative to SM of $0.607$, $0.543$, and $0.512$. Under dual misspecification, the corresponding borrowing probabilities were $30.9\%$, $4.4\%$, and $0.1\%$, while bias-detection probabilities were $68.9\%$, $95.6\%$, and $99.9\%$. The MSE ratios were $1.860$, $1.274$, and $1.007$. Hence the rule exhibits the predicted limiting behavior, but at $n=200$ it can still borrow too often under a fixed bias. The alternative $\{\log\log(n+e^e)\}^{1/2}$ threshold is more conservative in small samples: its dual-misspecification MSE ratios were $1.117$, $1.004$, and $1.000$, at the cost of smaller efficiency gains when the models were correct.

\subsection{Computational Implementation}

The Mac implementation parallelizes Monte Carlo batches with \texttt{mclapply}, while the Windows implementation uses a parallel socket (PSOCK) cluster; both default to the detected logical-core count minus one. Analytic gradients, warm starts, batch-level checkpointing, and bootstrapping only the S1EF estimators substantially reduce computation. The entropy-family choice is held fixed during the variance bootstrap, whereas the final MSE comparison is performed after the paired bootstrap quantities have been estimated.
\endgroup

\clearpage
For the real-data analysis, Table~\ref{supp:tab:summary} records the baseline summaries for the internal sample and the two external-data scenarios; Table~\ref{supp:tab:real-mse-selection} reports the MSE-based final decisions.

\begin{table}[htbp]
\centering
\caption{%
Summary information for the internal sample $(n=200)$ and the two external samples $(n=1{,}000$ each$)$, with absolute standardized mean differences (SMDs). ``SD'' denotes standard deviation, ``CPR'' cardiopulmonary resuscitation, and ``AED'' automated external defibrillator.
}
\vspace{1ex}
\label{supp:tab:summary}
\small
\setlength{\tabcolsep}{3pt}
\renewcommand{\arraystretch}{1.15}
\begin{tabular}{@{}p{0.32\textwidth}ccccc@{}}
\toprule
 & \shortstack{Internal\\$(n=200)$}
 & \multicolumn{2}{c}{\shortstack{External 2019\\$(n=1{,}000)$}}
 & \multicolumn{2}{c}{\shortstack{External 2010--15\\$(n=1{,}000)$}} \\
 & Summary & Summary & $|\mathrm{SMD}|$ & Summary & $|\mathrm{SMD}|$ \\ \midrule
Age (mean (SD))            & \twoline{64.89}{15.40} & \twoline{65.45}{15.99} & 0.036 & \twoline{65.35}{16.46} & 0.029 \\
Male (\%)                  & \twoline{156}{78.0}     & \twoline{820}{82.0}     & 0.100 & \twoline{793}{79.3}     & 0.032 \\
Family witness (\%)        & \twoline{99}{49.5}      & \twoline{477}{47.7}     & 0.036 & \twoline{494}{49.4}     & 0.002 \\
Bystander CPR (\%)         &                             &                             &       &                             &       \\
\quad none                 & \twoline{59}{29.5}      & \twoline{310}{31.0}     & 0.033 & \twoline{366}{36.6}     & 0.151 \\
\quad only chest compression & \twoline{111}{55.5}   & \twoline{583}{58.3}     & 0.057 & \twoline{480}{48.0}     & 0.151 \\
\quad with ventilation     & \twoline{30}{15.0}      & \twoline{107}{10.7}     & 0.129 & \twoline{154}{15.4}     & 0.011 \\
AED (\%)                   & \twoline{53}{26.5}      & \twoline{255}{25.5}     & 0.023 & \twoline{173}{17.3}     & 0.224 \\
Survival at 1 month (\%)   & \twoline{74}{37.0}      & \twoline{403}{40.3}     & 0.068 & \twoline{364}{36.4}     & 0.012 \\ \bottomrule
\end{tabular}
\end{table}

\begingroup
\begin{table}[htbp]
\centering
\caption{MSE-based final selection in the real-data analysis.}
\label{supp:tab:real-mse-selection}
\small
\setlength{\tabcolsep}{4pt}
\begin{tabular}{llrrrr}
\toprule
External & Method & Internal & Proposed & \shortstack{Estimated\\MSE ratio} & Decision \\ \midrule
2019 & EB & 0.370 & 0.399 & 0.422 & Proposed \\
2019 & EBW & 0.370 & 0.399 & 0.504 & Proposed \\
2010--15 & EB & 0.370 & 0.387 & 0.552 & Proposed \\
2010--15 & EBW & 0.370 & 0.387 & 0.507 & Proposed \\
\bottomrule
\end{tabular}
\end{table}
\endgroup

\clearpage
\begingroup
\setlength{\bibhang}{1.5em}
\setlength{\bibitemsep}{1pt}
\setlength{\emergencystretch}{2em}
\small
\printbibliography[title={References}]

@article{hu2022semiparametric,
  author  = {Hu, Wenjie and Wang, Ruoyu and Li, Wei and Miao, Wang},
  title   = {Semiparametric Efficient Fusion of Individual Data and Summary Statistics},
  journal = {J. Amer. Statist. Assoc.},
  year    = {2026},
  pages   = {1--28},
  doi     = {10.1080/01621459.2026.2659379},
  url     = {https://doi.org/10.1080/01621459.2026.2659379}
}

@article{rubin_inference_1976,
	title = {Inference and missing data},
	volume = {63},
	issn = {0006-3444, 1464-3510},
	url = {https://academic.oup.com/biomet/article-lookup/doi/10.1093/biomet/63.3.581},
	doi = {10.1093/biomet/63.3.581},
	language = {en},
	number = {3},
	urldate = {2024-10-08},
	journal = {Biometrika},
	author = {Rubin, Donald B.},
	year = {1976},
	pages = {581--592},
}

@article{pocock_combination_1976,
	title = {The combination of randomized and historical controls in clinical trials},
	volume = {29},
	copyright = {https://www.elsevier.com/tdm/userlicense/1.0/},
	issn = {00219681},
	url = {https://linkinghub.elsevier.com/retrieve/pii/0021968176900448},
	doi = {10.1016/0021-9681(76)90044-8},
	language = {en},
	number = {3},
	urldate = {2024-10-06},
	journal = {J. Chronic Dis.},
	author = {Pocock, Stuart J.},
	month = mar,
	year = {1976},
	pages = {175--188},
}

@book{borenstein_introduction_2009,
	edition = {1},
	title = {Introduction to {Meta}‐{Analysis}},
	copyright = {http://doi.wiley.com/10.1002/tdm\_license\_1.1},
	url = {https://onlinelibrary.wiley.com/doi/book/10.1002/9780470743386},
	language = {en},
	urldate = {2024-10-06},
	publisher = {Wiley},
	author = {Borenstein, Michael and Hedges, Larry V. and Higgins, Julian P. T. and Rothstein, Hannah R.},
	month = mar,
	year = {2009},
	doi = {10.1002/9780470743386},
}

@article{hainmueller_entropy_2012,
	title = {Entropy {Balancing} for {Causal} {Effects}: {A} {Multivariate} {Reweighting} {Method} to {Produce} {Balanced} {Samples} in {Observational} {Studies}},
	volume = {20},
	copyright = {https://www.cambridge.org/core/terms},
	issn = {1047-1987, 1476-4989},
	shorttitle = {Entropy {Balancing} for {Causal} {Effects}},
	url = {https://www.cambridge.org/core/product/identifier/S1047198700012997/type/journal_article},
	doi = {10.1093/pan/mpr025},
	language = {en},
	number = {1},
	urldate = {2024-10-06},
	journal = {Polit. Anal.},
	author = {Hainmueller, Jens},
	year = {2012},
	pages = {25--46},
}

@article{kitamura_nationwide_2010,
	title = {Nationwide {Public}-{Access} {Defibrillation} in {Japan}},
	volume = {362},
	issn = {0028-4793, 1533-4406},
	url = {http://www.nejm.org/doi/abs/10.1056/NEJMoa0906644},
	doi = {10.1056/NEJMoa0906644},
	language = {en},
	number = {11},
	urldate = {2024-10-02},
	journal = {New Engl. J. Med.},
	author = {Kitamura, Tetsuhisa and Iwami, Taku and Kawamura, Takashi and Nagao, Ken and Tanaka, Hideharu and Hiraide, Atsushi},
	month = mar,
	year = {2010},
	pages = {994--1004},
}

@article{josey_transporting_2021,
	title = {Transporting experimental results with entropy balancing},
	volume = {40},
	issn = {0277-6715, 1097-0258},
	url = {https://onlinelibrary.wiley.com/doi/10.1002/sim.9031},
	doi = {10.1002/sim.9031},
	language = {en},
	number = {19},
	urldate = {2024-09-30},
	journal = {Statist. Med.},
	author = {Josey, Kevin P. and Berkowitz, Seth A. and Ghosh, Debashis and Raghavan, Sridharan},
	month = aug,
	year = {2021},
	pages = {4310--4326},
}

@article{han_empirical_2019,
  title={Empirical likelihood estimation using auxiliary summary information with different covariate distributions},
  author={Han, Peisong and Lawless, Jerald F},
  journal={Statist. Sinica},
  volume={29},
  number={3},
  pages={1321--1342},
  year={2019},
  publisher={JSTOR}
}

@article{chatterjee_constrained_2016,
	title = {Constrained {Maximum} {Likelihood} {Estimation} for {Model} {Calibration} {Using} {Summary}-{Level} {Information} {From} {External} {Big} {Data} {Sources}},
	volume = {111},
	issn = {0162-1459, 1537-274X},
	url = {https://www.tandfonline.com/doi/full/10.1080/01621459.2015.1123157},
	doi = {10.1080/01621459.2015.1123157},
	language = {en},
	number = {513},
	urldate = {2024-10-08},
	journal = {J. Amer. Statist. Assoc.},
	author = {Chatterjee, Nilanjan and Chen, Yi-Hau and Maas, Paige and Carroll, Raymond J.},
	month = jan,
	year = {2016},
	pages = {107--117},
}

@article{shimodaira_improving_2000,
	title = {Improving predictive inference under covariate shift by weighting the log-likelihood function},
	volume = {90},
	copyright = {https://www.elsevier.com/tdm/userlicense/1.0/},
	issn = {03783758},
	url = {https://linkinghub.elsevier.com/retrieve/pii/S0378375800001154},
	doi = {10.1016/S0378-3758(00)00115-4},
	language = {en},
	number = {2},
	urldate = {2024-10-06},
	journal = {J. Statist. Plann. Inference},
	author = {Shimodaira, Hidetoshi},
	month = oct,
	year = {2000},
	pages = {227--244},
}

@article{zhao_entropy_2017,
	title = {Entropy {Balancing} is {Doubly} {Robust}},
	volume = {5},
	copyright = {De Gruyter expressly reserves the right to use all content for commercial text and data mining within the meaning of Section 44b of the German Copyright Act.},
	issn = {2193-3685},
	url = {https://www.degruyter.com/document/doi/10.1515/jci-2016-0010/html?lang=en},
	doi = {10.1515/jci-2016-0010},
	language = {en},
	number = {1},
	urldate = {2024-10-06},
	journal = {J. Causal Inference},
	author = {Zhao, Qingyuan and Percival, Daniel},
	month = mar,
	year = {2017},
}

@article{chu_targeted_2023,
	title = {Targeted optimal treatment regime learning using summary statistics},
	volume = {110},
	issn = {1464-3510},
	url = {https://doi.org/10.1093/biomet/asad020},
	doi = {10.1093/biomet/asad020},
	number = {4},
	urldate = {2024-10-06},
	journal = {Biometrika},
	author = {Chu, J and Lu, W and Yang, S},
	month = dec,
	year = {2023},
	pages = {913--931},
}

@article{qin_empirical_1994,
	title = {Empirical {Likelihood} and {General} {Estimating} {Equations}},
	volume = {22},
	issn = {0090-5364},
	url = {https://projecteuclid.org/journals/annals-of-statistics/volume-22/issue-1/Empirical-Likelihood-and-General-Estimating-Equations/10.1214/aos/1176325370.full},
	doi = {10.1214/aos/1176325370},
	language = {en},
	number = {1},
	urldate = {2024-10-24},
	journal = {Ann. Statist.},
	author = {Qin, Jin and Lawless, Jerry},
	month = mar,
	year = {1994},
	pages = {300--325},
}

@article{riley_meta-analysis_2010,
	title = {Meta-analysis of individual participant data: rationale, conduct, and reporting},
	volume = {340},
	issn = {0959-8138, 1468-5833},
	shorttitle = {Meta-analysis of individual participant data},
	url = {https://www.bmj.com/lookup/doi/10.1136/bmj.c221},
	doi = {10.1136/bmj.c221},
	language = {en},
	number = {feb05 1},
	urldate = {2025-03-18},
	journal = {BMJ},
	author = {Riley, R. D. and Lambert, P. C. and Abo-Zaid, G.},
	month = aug,
	year = {2010},
	pages = {c221},
}

@book{riley_individual_2021,
	title = {Individual {Participant} {Data} {Meta}-{Analysis} for {Healthcare} {Research}},
	isbn = {978-1-119-33378-4},
	url = {https://onlinelibrary.wiley.com/doi/abs/10.1002/9781119333784.ch1},
	language = {en},
	urldate = {2025-03-19},
	publisher = {John Wiley \& Sons, Ltd},
	author = {Riley, Richard D. and Stewart, Lesley A. and Tierney, Jayne F.},
	year = {2021},
	doi = {10.1002/9781119333784.ch1},
}

@article{lee_estimation_2020,
	title = {Estimation of {COVID}-19 spread curves integrating global data and borrowing information},
	volume = {15},
	issn = {1932-6203},
	url = {https://www.ncbi.nlm.nih.gov/pmc/articles/PMC7390340/},
	doi = {10.1371/journal.pone.0236860},
	number = {7},
	urldate = {2025-03-19},
	journal = {PLoS ONE},
	author = {Lee, Se Yoon and Lei, Bowen and Mallick, Bani},
	month = jul,
	year = {2020},
	pmid = {32726361},
	pmcid = {PMC7390340},
	pages = {e0236860},
}

@article{li_target_2020,
	title = {Target {Population} {Statistical} {Inference} {With} {Data} {Integration} {Across} {Multiple} {Sources}—{An} {Approach} to {Mitigate} {Information} {Shortage} in {Rare} {Disease} {Clinical} {Trials}},
	volume = {12},
	url = {https://doi.org/10.1080/19466315.2019.1654913},
	doi = {10.1080/19466315.2019.1654913},
	number = {3},
	urldate = {2025-03-19},
	journal = {Statist. Biopharm. Res.},
	author = {Li, Xihao and Song, Yang},
	month = jul,
	year = {2020},
	pages = {322--333},
}

@article{stewart_meta-analysis_1993,
	title = {Meta-analysis of the literature or of individual patient data: is there a difference?},
	volume = {341},
	copyright = {https://www.elsevier.com/tdm/userlicense/1.0/},
	issn = {01406736},
	shorttitle = {Meta-analysis of the literature or of individual patient data},
	url = {https://linkinghub.elsevier.com/retrieve/pii/014067369393004K},
	doi = {10.1016/0140-6736(93)93004-K},
	language = {en},
	number = {8842},
	urldate = {2025-03-19},
	journal = {Lancet},
	author = {Stewart, L.A and Parmar, M.K.B},
	month = feb,
	year = {1993},
	pages = {418--422},
}

@article{schafer_missing_2002,
	title = {Missing data: {Our} view of the state of the art},
	volume = {7},
	issn = {1939-1463, 1082-989X},
	shorttitle = {Missing data},
	url = {https://doi.apa.org/doi/10.1037/1082-989X.7.2.147},
	doi = {10.1037/1082-989X.7.2.147},
	language = {en},
	number = {2},
	urldate = {2025-03-19},
	journal = {Psychol. Methods},
	author = {Schafer, Joseph L. and Graham, John W.},
	year = {2002},
	pages = {147--177},
}

@book{little_statistical_2019,
	title = {Statistical {Analysis} with {Missing} {Data}},
	isbn = {978-0-470-52679-8},
	language = {en},
	publisher = {John Wiley \& Sons},
	author = {Little, Roderick J. A. and Rubin, Donald B.},
	month = apr,
	year = {2019},
	edition = {3},
}

@article{yang_combining_2020,
	title = {Combining {Multiple} {Observational} {Data} {Sources} to {Estimate} {Causal} {Effects}},
	volume = {115},
	issn = {0162-1459},
	url = {https://www.ncbi.nlm.nih.gov/pmc/articles/PMC7571608/},
	doi = {10.1080/01621459.2019.1609973},
	number = {531},
	urldate = {2025-03-19},
	journal = {J. Amer. Statist. Assoc.},
	author = {Yang, Shu and Ding, Peng},
	year = {2020},
	pmid = {33088006},
	pmcid = {PMC7571608},
	pages = {1540--1554},
}

@article{kundu_generalized_2019,
	title = {Generalized meta-analysis for multiple regression models across studies with disparate covariate information},
	volume = {106},
	issn = {0006-3444},
	url = {https://www.ncbi.nlm.nih.gov/pmc/articles/PMC6690173/},
	doi = {10.1093/biomet/asz030},
	number = {3},
	urldate = {2025-03-19},
	journal = {Biometrika},
	author = {Kundu, Prosenjit and Tang, Runlong and Chatterjee, Nilanjan},
	year = {2019},
	pmid = {31427822},
	pmcid = {PMC6690173},
	pages = {567--585},
}

@article{viele_use_2014,
	title = {Use of historical control data for assessing treatment effects in clinical trials},
	volume = {13},
	issn = {1539-1604},
	url = {https://www.ncbi.nlm.nih.gov/pmc/articles/PMC3951812/},
	doi = {10.1002/pst.1589},
	number = {1},
	urldate = {2025-03-19},
	journal = {Pharm. Statist.},
	author = {Viele, Kert and Berry, Scott and Neuenschwander, Beat and Amzal, Billy and Chen, Fang and Enas, Nathan and Hobbs, Brian and Ibrahim, Joseph G. and Kinnersley, Nelson and Lindborg, Stacy and Micallef, Sandrine and Roychoudhury, Satrajit and Thompson, Laura},
	year = {2014},
	pmid = {23913901},
	pmcid = {PMC3951812},
	pages = {41--54},
}

@article{qin_combining_2000,
	title = {Combining {Parametric} and {Empirical} {Likelihoods}},
	volume = {87},
	issn = {0006-3444},
	url = {https://www.jstor.org/stable/2673478},
	number = {2},
	urldate = {2025-03-19},
	journal = {Biometrika},
	author = {Qin, Jing},
	year = {2000},
	pages = {484--490},
}

@article{qin_using_2015,
	title = {Using covariate-specific disease prevalence information to increase the power of case-control studies},
	volume = {102},
	issn = {0006-3444},
	url = {https://www.jstor.org/stable/43305644},
	number = {1},
	urldate = {2025-03-19},
	journal = {Biometrika},
	author = {Qin, Jing and Zhang, Han and Li, Pengfei and Albanes, Demetrius and Yu, Kai},
	year = {2015},
	pages = {169--180},
}

@article{lumley_connections_2011,
	title = {Connections between {Survey} {Calibration} {Estimators} and {Semiparametric} {Models} for {Incomplete} {Data}},
	volume = {79},
	copyright = {© 2011 The Authors. Int. Statist. Rev. © 2011 International Statistical Institute},
	issn = {1751-5823},
	url = {https://onlinelibrary.wiley.com/doi/abs/10.1111/j.1751-5823.2011.00138.x},
	doi = {10.1111/j.1751-5823.2011.00138.x},
	language = {en},
	number = {2},
	urldate = {2025-03-19},
	journal = {Int. Statist. Rev.},
	author = {Lumley, Thomas and Shaw, Pamela A. and Dai, James Y.},
	year = {2011},
	pages = {200--220},
}

@article{cole_generalizing_2010,
	title = {Generalizing {Evidence} {From} {Randomized} {Clinical} {Trials} to {Target} {Populations}},
	volume = {172},
	issn = {0002-9262},
	url = {https://www.ncbi.nlm.nih.gov/pmc/articles/PMC2915476/},
	doi = {10.1093/aje/kwq084},
	number = {1},
	urldate = {2025-03-19},
	journal = {Amer. J. Epidemiol.},
	author = {Cole, Stephen R. and Stuart, Elizabeth A.},
	month = jul,
	year = {2010},
	pmid = {20547574},
	pmcid = {PMC2915476},
	pages = {107--115},
}

@article{jacobs_cardiac_2004,
	title = {Cardiac arrest and cardiopulmonary resuscitation outcome reports: update and simplification of the {Utstein} templates for resuscitation registries.: {A} statement for healthcare professionals from a task force of the international liaison committee on resuscitation ({American} {Heart} {Association}, {European} {Resuscitation} {Council}, {Australian} {Resuscitation} {Council}, {New} {Zealand} {Resuscitation} {Council}, {Heart} and {Stroke} {Foundation} of {Canada}, {InterAmerican} {Heart} {Foundation}, {Resuscitation} {Council} of {Southern} {Africa})},
	volume = {63},
	issn = {0300-9572, 1873-1570},
	shorttitle = {Cardiac arrest and cardiopulmonary resuscitation outcome reports},
	url = {https://www.resuscitationjournal.com/article/S0300-9572(04)00382-X/fulltext},
	doi = {10.1016/j.resuscitation.2004.09.008},
	language = {English},
	number = {3},
	urldate = {2025-03-19},
	journal = {Resuscitation},
	author = {Jacobs, Ian and Nadkarni, Vinay and Bahr, Jan and Berg, Robert A. and Billi, John E. and Bossaert, Leo and Cassan, Pascal and Coovadia, Ashraf and D’Este, Kate and Finn, Judith and Halperin, Henry and Handley, Anthony and Herlitz, Johan and Hickey, Robert and Idris, Ahamed and Kloeck, Walter and Larkin, Gregory Luke and Mancini, Mary Elizabeth and Mason, Pip and Mears, Gregory and Monsieurs, Koenraad and Montgomery, William and Morley, Peter and Nichol, Graham and Nolan, Jerry and Okada, Kazuo and Perlman, Jeffrey and Shuster, Michael and Steen, Petter Andreas and Sterz, Fritz and Tibballs, James and Timerman, Sergio and Truitt, Tanya and Zideman, David},
	month = dec,
	year = {2004},
	pages = {233--249},
}

@article{signorovitchComparativeEffectivenessHeadtoHead2010,
	title = {Comparative {Effectiveness} {Without} {Head}-to-{Head} {Trials}},
	volume = {28},
	issn = {1179-2027},
	url = {https://doi.org/10.2165/11538370-000000000-00000},
	doi = {10.2165/11538370-000000000-00000},
	language = {en},
	number = {10},
	urldate = {2025-06-06},
	journal = {PharmacoEconomics},
	author = {Signorovitch, James E. and Wu, Eric Q. and Yu, Andrew P. and Gerrits, Charles M. and Kantor, Evan and Bao, Yanjun and Gupta, Shiraz R. and Mulani, Parvez M.},
	month = oct,
	year = {2010},
	pages = {935--945},
}

@article{TaylorChoiHan_Biometrika_2023,
  author  = {Jeremy M. G. Taylor and Kyunghee Choi and Peisong Han},
  title   = {Data integration: exploiting ratios of parameter estimates from a reduced external model},
  journal = {Biometrika},
  year    = {2023},
  volume  = {110},
  number  = {1},
  pages   = {119--134},
  doi     = {10.1093/biomet/asac022}
}

@article{ZhaiHan_EJS_2024,
  author  = {Yuqi Zhai and Peisong Han},
  title   = {Integrating external summary information under population heterogeneity and information uncertainty},
  journal = {Electron. J. Statist.},
  year    = {2024},
  volume  = {18},
  pages   = {5304--5329},
  doi     = {10.1214/24-EJS2327}
}

@article{HanEtAl_Biometrics_2024,
  author  = {Peisong Han and Haoyue Li and Sung Kyun Park and Bhramar Mukherjee and Jeremy M. G. Taylor},
  title   = {Improving prediction of linear regression models by integrating external information from heterogeneous populations: {James--Stein} estimators},
  journal = {Biometrics},
  year    = {2024},
  volume  = {80},
  number  = {3},
  pages   = {ujae072},
  doi     = {10.1093/biomtc/ujae072}
}

@misc{bashari2025statistical,
  title         = {Statistical Inference Leveraging Synthetic Data with Distribution-Free Guarantees},
  author        = {Bashari, Meshi and Lee, Yonghoon and Lotan, Roy Maor and Dobriban, Edgar and Romano, Yaniv},
  year          = {2025},
  howpublished = {arXiv preprint arXiv:2509.20345},
  eprint        = {2509.20345},
  archivePrefix = {arXiv},
  primaryClass  = {stat.ME},
  doi           = {10.48550/arXiv.2509.20345},
  url           = {https://arxiv.org/abs/2509.20345},
  note          = {arXiv:2509.20345 [stat.ME]}
}

@article{robins_regression_1994,
  title        = {Estimation of regression coefficients when some regressors are not always observed},
  author       = {Robins, James M. and Rotnitzky, Andrea and Zhao, Lue Ping},
  journal      = {J. Amer. Statist. Assoc.},
  year         = {1994},
  volume       = {89},
  number       = {427},
  pages        = {846--866},
  doi          = {10.1080/01621459.1994.10476818}
}

@article{huang_simultaneous_2023,
	title = {Simultaneous selection and incorporation of consistent external aggregate information},
	volume = {42},
	issn = {0277-6715},
	doi = {10.1002/sim.9929},
	number = {30},
	journal = {Statist. Med.},
	author = {Huang, Yunxiang and Huang, Chiung-Yu and Kim, Mi-Ok},
	year = {2023},
	pages = {5630--5645}
}

@article{fang_integrated_2025,
	title = {An Integrated GMM Shrinkage Approach with Consistent Moment Selection from Multiple External Sources},
	volume = {34},
	number = {4},
	doi = {10.1080/10618600.2025.2476087},
	journal = {J. Comput. Graph. Statist.},
	author = {Fang, Fang and Long, Tian and Shao, Jun and Wang, Lei},
	year = {2025},
	pages = {1670--1679}
}
\endgroup
\end{document}